\documentclass[12pt, a4paper]{article}
\usepackage{cite}
\usepackage{amsmath,amssymb}
\input{colordvi.tex}
\usepackage{comment}
\usepackage{bm}
\usepackage{url}
\usepackage{braket}

\usepackage{ifpdf}
\ifpdf
  \usepackage{graphicx, hyperref, xcolor}     
\else     
  \usepackage[dvipdfmx]{graphicx, hyperref, xcolor}     
 \fi

\definecolor{rossoferrari}{HTML}{D9073D}
\definecolor{mediumblue}{HTML}{0000CD}
\hypersetup{
setpagesize=false,
bookmarksnumbered=true,%
bookmarksopen=true,%
colorlinks=true,%
linkcolor=rossoferrari,
urlcolor=mediumblue,
citecolor=mediumblue,
}

\begin{document}

\begin{titlepage}

\begin{center}

\hfill KEK-TH-2299\\

\vskip .75in

{\Large \bf
Axion/Hidden-Photon Dark Matter Conversion into \\[.3em] Condensed Matter Axion
}

\vskip .75in

{\large
So Chigusa$^{(a,b,c)}$, Takeo Moroi$^{(d,e)}$ and Kazunori Nakayama$^{(d,e)}$
}

\vskip 0.25in

$^{(a)}${\em Berkeley Center for Theoretical Physics, Department of Physics,\\
University of California, Berkeley, CA 94720, USA}\\[.3em]
$^{(b)}${\em Theoretical Physics Group, Lawrence Berkeley National Laboratory,\\
Berkeley, CA 94720, USA}\\[.3em]
$^{(c)}${\em KEK Theory Center, IPNS, KEK, Tsukuba, Ibaraki 305-0801, Japan}\\[.3em]
$^{(d)}${\em Department of Physics, Faculty of Science,\\
The University of Tokyo,  Bunkyo-ku, Tokyo 113-0033, Japan}\\[.3em]
$^{(e)}${\em Kavli IPMU (WPI), The University of Tokyo,  Kashiwa, Chiba 277-8583, Japan}

\end{center}
\vskip .5in

\begin{abstract}
The QCD axion or axion-like particles are candidates of dark matter of the universe. On the other hand, axion-like excitations exist in certain condensed matter systems, which implies that there can be interactions of dark matter particles with condensed matter axions. We discuss the relationship between the condensed matter axion and a collective spin-wave excitation in an anti-ferromagnetic insulator at the quantum level. The conversion rate of the light dark matter, such as the elementary particle axion or hidden photon, into the condensed matter axion is estimated for the discovery of the dark matter signals.
\end{abstract}

\end{titlepage}


\renewcommand{\thepage}{\arabic{page}}
\setcounter{page}{1}
\renewcommand{\thefootnote}{\#\arabic{footnote}}
\setcounter{footnote}{0}
\renewcommand{\theequation}{\thesection.\arabic{equation}}


\newpage

\tableofcontents

\newpage

\section{Introduction}
\label{sec:Intro}
\setcounter{equation}{0}

The QCD axion is a hypothetical elementary particle that solves the strong CP problem~\cite{Peccei:1977hh,Weinberg:1977ma,Wilczek:1977pj} and is a candidate of dark matter (DM) of the universe~\cite{Preskill:1982cy,Abbott:1982af,Dine:1982ah} (see Refs.~\cite{Kim:1986ax,Kim:2008hd,Kawasaki:2013ae} for reviews).
Recently people often consider axion-like particles (ALPs) in a broad sense, partly motivated by the developments in string theory~\cite{Svrcek:2006yi,Arvanitaki:2009fg,Cicoli:2012sz}. ALPs do not necessarily address the strong CP problem, but they are also good DM candidates and may be experimentally probed through, e.g., the axion-photon coupling of the form $\mathcal L\propto a \vec E\cdot\vec B$ where $a$ denotes the ALP field and $\vec E$ $(\vec B)$ denotes the electric (magnetic) field respectively. There are many experimental ideas to search for ALPs including the QCD axion,\footnote{
	In the following we use the terminology ``ALP'' for general elementary axion-like particles including the QCD axion.
} although still it is not discovered yet~\cite{Sikivie:1983ip,Bradley:2003kg,Asztalos:2009yp,Zhong:2018rsr,McAllister:2017lkb,Alesini:2017ifp,Semertzidis:2019gkj,Horns:2012jf,Jaeckel:2013eha,TheMADMAXWorkingGroup:2016hpc,Kahn:2016aff,Obata:2018vvr,Nagano:2019rbw,Lawson:2019brd,Zarei:2019sva,Budker:2013hfa,Barbieri:1985cp,Barbieri:2016vwg,Crescini:2020cvl,Chigusa:2020gfs,Marsh:2018dlj}.

On the other hand, the axion-like excitation also appears in the condensed matter physics~\cite{Wilczek:1987mv,Li:2009tca} (see Refs.~\cite{Sekine:2020ixs,Nenno:2020ujq} for reviews). To distinguish it from the elementary particle axion or ALP, we call such an axion-like excitation in condensed matter context as ``condensed matter axion (CM axion)''. The CM axion $\delta\theta$ has an interaction with the electromagnetic field as $\mathcal L \propto \delta\theta \vec E\cdot\vec B$, similar to the ALP. We call such an insulator an axionic insulator.

Let us briefly mention a relation between the topological insulator and axionic insulator. In general, topological electromagnetic responses of a three-dimensional insulator are described by the topological term in the Lagrangian:
\begin{align}
	\mathcal L = \theta\frac{\alpha_e}{8\pi}F_{\mu\nu}\widetilde F^{\mu\nu}
	=\theta\frac{\alpha_e}{2\pi}\vec E\cdot\vec B.
\end{align}
For example, it implies that there appears a magnetization (electric polarization) proportional to the applied electric (magnetic) field: $\vec M\propto \theta\vec E$ $(\vec P\propto \theta \vec B)$. If the Hamiltonian of the system is invariant under the time-reversal symmetry, the coefficient $\theta$ can only take a value either $0$ or $1/2$: i.e., such an insulator is classified by a discrete $Z_2$ index~\cite{Kane:2005zz,Fu:2007uya,Fu:2007,Hasan:2010xy,Qi:2011zya,Bernevig:1529190}.\footnote{
	Time-reversal invariant topological insulators have been first considered in two-dimensional systems~\cite{Kane:2004bvs,Bernevig_2006}.
} The case of $\theta=1/2$ corresponds to the topological insulator, in which the existence of gapless surface states is ensured and it causes topological electromagnetic effects.
On the other hand, if there is no time-reversal symmetry, $\theta$ does not have to be quantized but can take arbitrary values possibly with a space-time dependence: $\theta=\theta(\vec x,t)$. If $\theta$ is a dynamical field, it is called the CM axion. Although it is often helpful to start from the topological insulator for understanding the origin of CM axion, the existence of CM axion does not necessarily require that the insulator is topological. One can generally write $\theta(\vec x,t)=\theta_0+\delta\theta(\vec x,t)$ so that $\delta\theta(\vec x,t)$ expresses the CM axion while $\theta_0$ is the background value.  
The value of $\theta_0$ depends on the properties of the material and can be zero. 
It has been known that in a class of magnetically doped topological insulators, the fluctuation of the anti-ferromagnetic order parameter (the so-called Neel field) plays a role of CM axion~\cite{Li:2009tca}.

In this paper, we consider a process like the light DM conversion into the CM axion and estimate the conversion rate. Such a process has been considered in Ref.~\cite{Marsh:2018dlj} for the detection of axion-like DM.
One of the main purposes of this paper is to discuss the origin of CM axion in a comprehensive and self-consistent manner for particle physicists. We will explicitly show the relationship between the CM axion and the spin-wave fluctuation (magnon) based on a model presented in Ref.~\cite{Sekine:2014xva}.
Another purpose is to provide a useful method to calculate the DM conversion rate into the CM axion in a quantum mechanical way. As an illustration, we will consider the case of ALP DM and hidden-photon DM.

This paper is organized as follows. In Sec.~\ref{sec:Heisenberg} we review the (anti-ferromagnetic) Heisenberg model of the localized electron spin system on the lattice. It gives a basis of the collective spin-wave excitation (magnon) and its dispersion relation, which will turn out to be identified with the CM axion in a certain setup. In Sec.~\ref{sec:Hubbard} the so-called (half-filling) Hubbard model is briefly introduced. Electrons in solids are often modeled by a tight-binding Hamiltonian plus the Coulomb repulsive force between electrons on the same lattice point (Hubbard interaction). It is shown that the limit of large Hubbard interaction reduces to the (anti-ferromagnetic) Heisenberg model. Therefore, the Hubbard model on a certain lattice may describe both the electron energy band structure as well as the anti-ferromagnetic order and magnon excitation around it.
In Sec.~\ref{sec:CM} we introduce the Fu-Kane-Mele-Hubbard model as a concrete setup and show that it contains an excitation that is regarded as the CM axion along the line of Ref.~\cite{Sekine:2014xva}. It will become clear that the CM axion is described by the use of anti-ferromagnetic magnon and its dispersion can be estimated as explained in Sec.~\ref{sec:Heisenberg}.
In Sec.~\ref{sec:DM} we estimate the conversion rate of light bosonic DM into the CM axion. We consider two DM models: ALP and hidden photon.  We conclude in Sec.~\ref{sec:conc}.

\section{Magnon in anti-ferromagnet}
\label{sec:Heisenberg}
\setcounter{equation}{0}

Let us start with the Heisenberg anti-ferromagnet model~\cite{Kittel:1951a,Kittel:1951b,Quinn:2009}.\footnote{As explained in Sec.~\ref{sec:Hubbard}, the Heisenberg anti-ferromagnet model may be understood from the Hubbard model in the limit of strong electron self-interaction at each site.}
Suppose a bipartite lattice consisting of sublattices A and B, and on each lattice point $\ell\in$ A or $\ell'\in$ B there is an electron spin $\vec S$. Applying an external magnetic field $B_0$ along the $z$ direction, the model Hamiltonian is given by
\begin{align}
	H = -\frac{J}{2}\sum_{\left<\ell,\ell'\right>}\vec S_\ell\cdot \vec S_{\ell'}
	-g\mu_B (B_A+B_0)\sum_{\ell\in A} S_{\ell}^z+g\mu_B (B_A-B_0)\sum_{\ell'\in B}S_{\ell'}^z,
	\label{Heisenberg}
\end{align}
where $J<0$ is the exchange interaction, $g=2$ and $\mu_B=e/(2m_e)$ is the Bohr magneton, and $B_A$ is the anisotropy field. 
The collective excitation of the spin-wave around the ground state, called magnon, is analyzed through the Holstein-Primakoff transformation,
\begin{align}
	&S_{\ell}^+ = \sqrt{2s-a_\ell^\dagger a_\ell}\,a_\ell,~~~
	S_{\ell}^- = a_\ell^\dagger\sqrt{2s-a_\ell^\dagger a_\ell},~~~
	S_\ell^z = s- a_\ell^\dagger a_\ell, \\
	&S_{\ell'}^+ =b_{\ell'}^\dagger\sqrt{2s-b_{\ell'}^\dagger b_{\ell'}},~~~
	S_{\ell'}^- = \sqrt{2s-b_{\ell'}^\dagger b_{\ell'}}\,b_{\ell'},~~~
	S_{\ell'}^z = -s +b_{\ell'}^\dagger b_{\ell'},
\end{align}
where we have defined $S_{\ell}^\pm=S_\ell^x\pm iS_{\ell}^y$ and $S_{\ell'}^\pm=S_{\ell'}^x\pm iS_{\ell'}^y$, and the creation-annihilation operators satisfy the commutation relation
\begin{align}
	\left[ a_{\ell}, a_{m}^\dagger \right] = \delta_{\ell m},~~~~~~\left[ b_{\ell'}, b_{m'}^\dagger \right] = \delta_{\ell' m'}.
\end{align}
In addition, $s$ is the spin quantum number; the eigenvalue of $\vec{S}_\ell\cdot\vec{S}_\ell$ is given by $s(s+1)$.
The Hamiltonian is rewritten in terms of the creation-annihilation operators as
\begin{align}
	H =& \, 2Nzs^2 J -2Ns \omega_A - Js \sum_{\left<\ell,\ell'\right>}
	\left( a_{\ell}^\dagger a_{\ell} + b_{\ell'}^\dagger b_{\ell'} +  a_{\ell}^\dagger b_{\ell'}^\dagger+  a_{\ell} b_{\ell'} \right) \nonumber\\
	&+(\omega_A+\omega_L)\sum_\ell a_{\ell}^\dagger a_{\ell}
	+(\omega_A-\omega_L)\sum_{\ell'} b_{\ell'}^\dagger b_{\ell'},
\end{align}
where $N$ is the total number of sites in a sublattice, $z$ denotes the number of adjacent lattice points (e.g. $z=6$ for simple bipartite cubic lattice), and
\begin{align}
	\omega_L\equiv g\mu_B B_0,~~~~~~\omega_A\equiv g\mu_B B_A.
\end{align}

Now let us move to the Fourier space. We define the Fourier component as
\begin{align}
	a_\ell = \frac{1}{\sqrt N}\sum_{\vec k} e^{-i\vec k\cdot \vec x_\ell} a_{\vec k},~~~~~~
	b_{\ell'} = \frac{1}{\sqrt N}\sum_{\vec k} e^{i\vec k\cdot \vec x_{\ell'}} b_{\vec k}.
\end{align}
Substituting this into the Hamiltonian, we find
\begin{align}
	H=\sum_{\vec k}\left[  (\omega_J+\omega_A+\omega_L) a_{\vec k}^\dagger a_{\vec k}
	 +(\omega_J+\omega_A-\omega_L) b_{\vec k}^\dagger b_{\vec k}
	 +\omega_J \gamma_{\vec k} (a_{\vec k} b_{\vec k}+a_{\vec k}^\dagger b^\dagger_{\vec k}) \right],
\end{align}
where $\omega_J\equiv -2zs J$ and
\begin{align}
	\gamma_{\vec k}= \frac{1}{z}\sum_{\vec \delta} e^{i\vec k\cdot \vec \delta},
\end{align}
with $\vec \delta$ being the vector connecting the adjacent lattice points. Finally, it is diagonalized through the Bogoliubov transformation:
\begin{align}
	\alpha_{\vec k} = u_{\vec k} a_{\vec k}-v_{\vec k}b_{\vec k}^\dagger,~~~~~~
	\beta_{\vec k}^\dagger = u_{\vec k} b_{\vec k}^\dagger-v_{\vec k}a_{\vec k}.
\end{align}
One can check that the canonical commutation relation is maintained if $|u_{\vec k}|^2-|v_{\vec k}|^2=1$. The concrete expression is given by
\begin{align}
	|u_{\vec k}|^2=\frac{1}{2}\left(1+\frac{\omega_J+\omega_A}{\sqrt{(\omega_J+\omega_A)^2-|\gamma_{\vec k}|^2\omega_J^2}} \right),~~~~
	|v_{\vec k}|^2=\frac{1}{2}\left(-1+\frac{\omega_J+\omega_A}{\sqrt{(\omega_J+\omega_A)^2-|\gamma_{\vec k}|^2\omega_J^2}} \right),
	\label{eq:u2v2}
\end{align}
with $\arg (\gamma_{\vec{k}}) = 2\arg (u_{\vec{k}}) = -2\arg (-v_{\vec{k}})$.  (Thus, $u_{\vec{k}}v_{\vec{k}}$ is real and negative.) Note that, when $\omega_A\ll \omega_J$, we have large Bogoliubov coefficients $|u_{\vec k}|^2\sim |v_{\vec k}|^2\gg1$ for $|\vec k\cdot\vec \delta|\ll 1$.
Then, one finds the diagonal Hamiltonian:
\begin{align}
	H = \sum_{\vec k}\left[ (\omega_{\vec k}+\omega_L)\alpha_{\vec k}^\dagger\alpha_{\vec k}
	+ (\omega_{\vec k}-\omega_L)\beta_{\vec k}^\dagger\beta_{\vec k} \right].
  \label{eq:H0-magnon}
\end{align}
Here, $\omega_{\vec k}$ represents the magnon dispersion relation (besides the overall offset coming from the Larmor frequency $\omega_L$),
\begin{align}
	\omega_{\vec k}^2=\omega_J^2(1-|\gamma_{\vec k}|^2)+\omega_A (\omega_A+2\omega_J).
\end{align}
In the low frequency limit $|\vec k\cdot\vec \delta|\ll 1$, we obtain $\gamma_{\vec k}\simeq 1+i\sum_{\vec \delta}(\vec{k}\cdot\vec{\delta})/z -\sum_{\vec \delta}(\vec k\cdot\vec\delta)^2/z$.
It implies the linear dispersion relation, $\omega_{\vec k}\propto |\vec k|$ for large $|\vec k|$ (but still it satisfies $|\vec k\cdot\vec \delta|\ll 1$), in contrast to the ferromagnetic magnon dispersion relation, which would show $\omega_{\vec k} \propto k^2$.
They are related to the so-called type-I and type-II Nambu-Goldstone boson dispersion relation as generally classified in Refs.~\cite{Watanabe:2012hr,Hidaka:2012ym}.

\section{Hubbard model as origin of anti-ferromagnet}
\label{sec:Hubbard}
\setcounter{equation}{0}

\subsection{Tight-binding model}

A tight-binding model is one of the approaches to estimate the electron energy band structure in solids. In this approach, one starts with the picture that each electron is rather tightly bounded by each atom and then takes into account the overlap between the nearest electron wave function.

Let us consider only one electron orbital at each site and neglect the interaction among different orbits, spin-orbit coupling, electron self-interaction, etc.\footnote{
	Effects of the interaction among different orbitals and spin-orbit coupling are important for the topological insulator. The electron self-interaction will be taken into account in the next subsection.
}
In the second quantization picture, the tight-binding Hamiltonian is given by
\begin{align}
	H=-t\sum_{\left<i,j\right>,\sigma}c_{i\sigma}^\dagger c_{j\sigma},
\end{align}
where $c_{i\sigma}^\dagger$ and $c_{i\sigma}$ denote the electron creation and annihilation operators at the site $i$ with spin $\sigma$ ($\uparrow$ or $\downarrow$) and the summation is taken over the combination of adjacent sites $\left<i,j\right>$. The creation and annihilation operators satisfy the anti-commutation relation
\begin{align}
	\left\{ c_{i\sigma}, c^\dagger_{j\sigma'} \right\} = \delta_{ij}\delta_{\sigma\sigma'}.
\end{align}
The Fourier transformation is defined by
\begin{align}
	c_{i\sigma} = \frac{1}{\sqrt N}\sum_{\vec k} e^{-i\vec k\cdot \vec x_i} c_{\vec k,\sigma}.
\end{align}
The Hamiltonian is rewritten in a diagonal form as
\begin{align}
	H = \sum_{\vec k,\sigma} \epsilon_{\vec k} c_{\vec k,\sigma}^\dagger c_{\vec k,\sigma},
	~~~~~~\epsilon_{\vec k}=-t(\gamma_{\vec k}+\gamma_{\vec k}^*).
\end{align}
This $\epsilon_{\vec k}$ denotes the electron energy band. In a simple cubic lattice, for example, we obtain $\epsilon_k=2t\left(1-\sum_{i=x,y,z}\cos(k_i a)\right)$.

The conductivity of this model is determined by the number of electrons in the system. If each orbital is filled, i.e., there are two electrons with opposite spins at each site, the energy band is filled and this becomes an insulator as far as there is an energy gap to the next energy band. If there is only one electron at each orbital, the energy band is not filled and it becomes a metal.

\subsection{Half-filling Hubbard model}

Let us add the effect of interaction between electrons at the same site $i$ to the tight-binding Hamiltonian. The resulting Hamiltonian is called the Hubbard model:
\begin{align}
	H=H_{t}+H_{U}= -t\sum_{\left<i,j\right>,\sigma}c_{i\sigma}^\dagger c_{j\sigma}
	+ U \sum_i n_{i\uparrow} n_{i\downarrow},
        \label{eq:hubbard}
\end{align}
where $U>0$ represents the interaction energy and $n_{i\uparrow} = c_{i\uparrow}^\dagger c_{i\uparrow}$ and $n_{i\downarrow} = c_{i\downarrow}^\dagger c_{i\downarrow}$.

The Hubbard model is characterized by several parameters: the relative interaction strength $U/t$ and the number of electrons per site, $N_e/N_s$. The case of $N_e/N_s=1$ is called the half-filling (it is ``half'' because of the spin degree of freedom) and its properties are well understood. Below, we consider the half-filling case.
Naively, one may consider that the half-filling Hubbard model describes a metal since electrons are in a conducting band. It is true in the limit $U=0$, but it is not necessarily true for sizable interaction strength. The interaction term can split the energy band and make a gap, which would result in an insulator. Such an insulator is called the Mott insulator.

Now we consider the large interaction limit: $U/t\gg 1$. In this limit, the tight-binding part is regarded as a perturbation. In the ground state, one electron is localized at each site to minimize the Hubbard interaction energy (hence it is expected that it behaves as an insulator rather than metal). Thus, the ground state is expressed as
\begin{align}
	\left| \widetilde\sigma \right>=\left(\prod_{i} c_{i\sigma_i}^\dagger\right)\left| 0\right>,
	\label{Hubbard_ground}
\end{align}
where $\widetilde\sigma$ schematically represents the array of spin, e.g., $\widetilde\sigma=(\dots,\uparrow,\uparrow,\downarrow,\dots)$ and so on. There are $2^{N_e}$ degenerate ground states corresponding to the spin degree of freedom at each site.

We want to consider an effective Hamiltonian regarding $H_t$ as a perturbation. Noting $\left<\widetilde\sigma\right| H_{t} \left|\widetilde\sigma\right>=0$, the nontrivial effect appears at the second-order in $H_t$. The effective Hamiltonian is given by
\begin{align}
	H_{\rm eff } = - \mathcal P H_t \frac{1}{H_{U}} H_t \mathcal P 
	 = -\frac{t^2}{U}\mathcal P  \sum_{\left<i,j\right>\sigma\sigma'}\left(
	c_{i\sigma}^\dagger c_{j\sigma} c_{j\sigma'}^\dagger c_{i\sigma'}+
	c_{j\sigma}^\dagger c_{i\sigma} c_{i\sigma'}^\dagger c_{j\sigma'}
	\right)  \mathcal P,
\end{align}
where $\mathcal P$ denotes the projection operator to the Hilbert space spanned by the ground state (\ref{Hubbard_ground}). The physical meaning is that, for $\sigma\neq\sigma'$, it exchanges the spin at the adjacent sites $i$ and $j$ for a given ground state.
This is rewritten in terms of the spin operator as
\begin{align}
	H_{\rm eff} = \frac{4t^2}{U} \sum_{\left<i,j\right>}\vec S_i\cdot \vec S_j,
\end{align}
where we have defined
\begin{align}
	S_i^z=\frac{1}{2}(c_{i\uparrow}^\dagger c_{i\uparrow}-c_{i\downarrow}^\dagger c_{i\downarrow}),~~~~~
	S_i^+\equiv S_i^x+iS_y=c_{i\uparrow}^\dagger c_{i\downarrow},~~~~~S_i^-\equiv S_x-iS_y=c_{i\downarrow}^\dagger c_{i\uparrow}.
\end{align}
Since the coefficient $t^2/U$ is positive, it represents the Heisenberg anti-ferromagnet model with $J=-t^2/U$. Thus, the half-filling Hubbard model may describe both the metal phase in the limit $U\to 0$ and the anti-ferromagnetic insulator phase in the large $U$ limit.

\section{A model of condensed matter axion}
\label{sec:CM}
\setcounter{equation}{0}

\subsection{Energy band in Fu-Kane-Mele-Hubbard model}

A three-dimensional topological insulator has been proposed in Refs.~\cite{Fu:2007uya,Fu:2007}. An example is the diamond lattice with a strong spin-orbit coupling.
On the other hand, taking account of the Hubbard on-site interaction between electrons may lead to the anti-ferromagnetic phase, leading to the topological anti-ferromagnet.
Such a model is called the Fu-Kane-Mele-Hubbard model and studied in Ref.~\cite{Sekine:2014xva}.
Actually, it is found in Ref.~\cite{Sekine:2014xva} that there is a topological anti-ferromagnetic phase depending on the interaction strength, in which the spin-wave excitation (magnon) has an axionic coupling to the electromagnetic field.

Now, we briefly review the Fu-Kane-Mele-Hubbard model on the diamond
lattice. We assume the half-filling case, i.e., there is only one
electron at the electron orbitals of our interest at each site.
The model Hamiltonian is given by $H=H_0 +
H_U$:
\begin{align}
	&H_0 = \sum_{\left<i,j\right>\sigma} t_{ij} c_{i\sigma}^\dagger c_{j\sigma}+
	i\frac{4\lambda}{a^2}\sum_{\left<\left<i,j\right>\right>} c_{i}^\dagger \vec\sigma\cdot(\vec d^1_{ij}\times \vec d_{ij}^2) c_j,
  \label{eq:TB-SO}\\
	&H_U = U \sum_i n_{i\uparrow} n_{i\downarrow},
\end{align}
where $c_i\equiv (c_{i\uparrow}, c_{i\downarrow})^T$.  Here, $\vec
d^1_{ij}$ and $\vec{d}^2_{ij}$ are the two vectors that connect two adjacent
sites: $\frac{a}{4}(1,1,1)$, $\frac{a}{4}(1,-1,-1)$,
$\frac{a}{4}(-1,1,-1)$, $\frac{a}{4}(-1,-1,1)$, with $a$ being the
lattice constant and $\lambda$ represents the strength of the
spin-orbit coupling. Note that the diamond lattice consists of two
sublattices (which we call A and B) both of which are face-centered
cubic. $\left<\left<i,j\right>\right>$ denotes a set of the
next-nearest neighbor sites, and hence sites $i$ and $j$ belong to the
same sublattice.  (For more detail about the interaction of electrons in next-nearest neighbor sites, see App.\ \ref{sec:spin}.)

Let us study the energy bands of this model neglecting the Hubbard interaction term~\cite{Fu:2007uya,Fu:2007}.
In the Fourier space, the Hamiltonian is expressed as the matrix form in the basis $c_{\vec k} \equiv (c_{\vec k\uparrow,A},c_{\vec k\downarrow,A},c_{\vec k\uparrow,B},c_{\vec k\downarrow,B})^T$ as
\begin{align}
	H_0 = \sum_{\vec k} c_{\vec k}^\dagger \mathcal H c_{\vec k},~~~~~
	\mathcal H=\sum_{\mu=1}^5 R_\mu(\vec k) \alpha_\mu,
  \label{eq:four-band}
\end{align}
where
\begin{align}
	&R_1(\vec k)=\lambda\left[\sin(\vec k\cdot\vec a_2)-\sin(\vec k\cdot\vec a_3)-\sin(\vec k\cdot(\vec a_2-\vec a_1))-\sin(\vec k\cdot(\vec a_3-\vec a_1))  \right],\\
	&R_2(\vec k)=\lambda\left[\sin(\vec k\cdot\vec a_3)-\sin(\vec k\cdot\vec a_1)-\sin(\vec k\cdot(\vec a_3-\vec a_2))-\sin(\vec k\cdot(\vec a_1-\vec a_2))  \right],\\
	&R_3(\vec k)=\lambda\left[\sin(\vec k\cdot\vec a_1)-\sin(\vec k\cdot\vec a_2)-\sin(\vec k\cdot(\vec a_1-\vec a_3))-\sin(\vec k\cdot(\vec a_2-\vec a_3))  \right],\\
	& R_4(\vec k)=t\left[ 1+\cos(\vec k\cdot\vec a_1)+\cos(\vec k\cdot\vec a_2)+\cos(\vec k\cdot\vec a_3) \right]+\delta t,\\
	& R_5(\vec k)=t\left[\sin(\vec k\cdot\vec a_1)+\sin(\vec k\cdot\vec a_2)+\sin(\vec k\cdot\vec a_3) \right],
\end{align}
with $\vec a_1=\frac{a}{2}(0,1,1), \vec a_2=\frac{a}{2}(1,0,1), \vec a_3=\frac{a}{2}(1,1,0)$ and
\begin{align}
	\alpha_i=\begin{pmatrix}
		\sigma_i & 0 \\ 0 &-\sigma_i
	\end{pmatrix},~~~
	\alpha_4=\begin{pmatrix}
		0 & 1 \\ 1 & 0
	\end{pmatrix},~~~
	\alpha_5=\begin{pmatrix}
		0 & i \\ -i & 0
	\end{pmatrix}.
\end{align}
These $\alpha$ matrices are Hermite and satisfy the anti-commutation relation $\{\alpha_\mu,\alpha_\nu\}=2\delta_{\mu\nu}$. Then, it is easy to show that the energy eigenvalues are given by
\begin{align}
	E_\pm
	= \pm \sqrt{\sum_{\mu} \left(R_\mu(\vec k)\right)^2}.
\end{align}
This gives the dispersion relation of the bulk electron.
It is found that, at the so-called $X_r$ points $(r=1,2,3)$ of the momentum space, $\vec k_{X_1}=\frac{2\pi}{a}(1,0,0), \vec k_{X_2}=\frac{2\pi}{a}(0,1,0), \vec k_{X_3}=\frac{2\pi}{a}(0,0,1)$, which are located at the boundary of the Brillouin zone, we obtain $E_\pm=0$ in the limit of $\delta t=0$. Thus, this material is regarded as a semimetal in this limit.
For example, the dispersion relation around $\vec k=\vec k_{X_1}$ is given by
\begin{align}
	E_{\pm}(\vec q)=\pm\sqrt{ (t q_x)^2 + 4\lambda^2(q_y^2+q_z^2) + (\delta t)^2},
\end{align}
where we have taken $\vec k= \vec k_{X_1} + \vec q$. Thus, nonzero $\delta t$ gives the energy gap between two energy bands, which makes the material the bulk insulator (topological insulator, actually).

\subsection{Axionic excitation in anti-ferromagnetic phase} \label{sec:cond-axion}

It is expected that the inclusion of the Hubbard interaction $H_U$ may lead to the anti-ferromagnetic ordering. Actually, it is found that the anti-ferromagnetic phase appears for sizable $U/t$ in the mean field approximation~\cite{Sekine:2014xva}.
Under this approximation, the Hubbard interaction term can be rewritten as
\begin{align}
  H_U \simeq U\sum_i & \Big( \Braket{n_{i\uparrow}} n_{i\downarrow}
  + \Braket{n_{i\downarrow}} n_{i\uparrow}
  - \Braket{n_{i\uparrow}} \Braket{n_{i\downarrow}} \notag \\
  &- \Braket{c_{i\uparrow}^\dagger c_{i\downarrow}} c_{i\downarrow}^\dagger c_{i\uparrow}
  - \Braket{c_{i\downarrow}^\dagger c_{i\uparrow}} c_{i\uparrow}^\dagger c_{i\downarrow}
  + \Braket{c_{i\uparrow}^\dagger c_{i\downarrow}} \Braket{c_{i\downarrow}^\dagger c_{i\uparrow}} \Big),
\end{align}
with $\Braket{\mathcal{O}}$ being the ensemble average of the operator $\mathcal{O}$.
We use the operator equations
\begin{align}
  n_{i\uparrow (\downarrow)} &= \pm S_i^{'z} + \frac{1}{2} (n_{i\uparrow} + n_{i\downarrow}), \label{eq:n_iud}\\
  c_{i\uparrow}^\dagger c_{i\downarrow} &= S_i^{'x} + iS_i^{'y},\\
  c_{i\downarrow}^\dagger c_{i\uparrow} &= S_i^{'x} - iS_i^{'y}, \label{eq:cdu}
\end{align}
with $\vec{S}_i'$ being spin operators in the coordinate system used in the previous subsection, with which three Dirac points are defined.
Note that, in the $U\to \infty$ limit of a half-filling model, we can safely restrict ourselves to states with $\Braket{n_{i\uparrow} + n_{i\downarrow}} = 1$.
Then, neglecting constant terms, the Hubbard interaction becomes
\begin{align}
  H_U \ni \sum_{\vec{k}} c_{\vec{k}}^\dagger \mathcal{H}_U c_{\vec{k}},~~~~~
  \mathcal{H}_U = -U \sum_{r=1}^3 m_r \alpha_r,
\end{align}
with $m_r$ are defined through
\begin{align}
	\left< \vec S_{i,A}\right>=-\left< \vec S_{i,B}\right> \equiv \vec m,
\end{align}
which characterizes the anti-ferromagnetic ordering.

Under this background and assuming $U|\vec m|\ll \lambda$, the $X_r$ points $(r=1,2,3)$ are slightly shifted as
\begin{align}
	\vec k_{\widetilde X_1}=\left(\frac{2\pi}{a},\frac{Um_2}{2\lambda a},-\frac{Um_3}{2\lambda a}\right),
	\vec k_{\widetilde X_2}=\left(-\frac{Um_1}{2\lambda a},\frac{2\pi}{a},\frac{Um_3}{2\lambda a}\right),
	\vec k_{\widetilde X_3}=\left(\frac{Um_1}{2\lambda a},-\frac{Um_2}{2\lambda a},\frac{2\pi}{a}\right).
\end{align}
For example, the energy dispersion around the $\widetilde X_1$ point is given by
\begin{align}
	E_{\pm}(\vec q)=\pm\sqrt{ (t q_x)^2 + 4\lambda^2(q_y^2+q_z^2) + (\delta t)^2 + (Um_1)^2},
\end{align}
where we have taken $\vec k= \vec k_{\widetilde X_1} + \vec q$. It is seen that there is an additional gap due to the anti-ferromagnetic order.

The Hamiltonian around the $\widetilde X_1$ point is expressed as
\begin{align}
	\mathcal H_{\widetilde X_1}(\vec q)=\frac{1}{a}(\widetilde q_x \alpha_1+\widetilde q_y \alpha_2+\widetilde q_z \alpha_3 )+\delta t\,\alpha_4+ Um_1 \alpha_5,
	\label{eq:HX1}
\end{align}
where we have rescaled the momentum as $t q_x \to \widetilde q_x/a$, $2\lambda q_y \to \widetilde q_y/a$, $2\lambda q_z \to \widetilde q_z/a$.
In deriving Eq.~\eqref{eq:HX1}, we have performed an appropriate change of the basis of the $\alpha$ matrices through a unitary transformation, with which $\alpha_1\leftrightarrow\alpha_5$
(see App.\ \ref{sec:alpha}).  The Hamiltonian around the $\widetilde X_2$ and $\widetilde X_3$ points can also be reduced to the same form except for the last term, which becomes $Um_2\alpha_5$ and $Um_3\alpha_5$, respectively.
From this Hamiltonian, we can infer the effective action for the electron which mimics the action of the relativistic Dirac fermion as
\begin{align}
	S=\int d^4x \sum_{r=1,2,3}\overline \psi_r \left[ i\gamma^\mu (\partial_\mu-ieA_\mu)-\delta t -i\gamma_5 Um_r  \right]\psi_r.
\end{align}

One can make a chiral rotation of the fermion to eliminate the $\gamma_5$ dependent term, $\psi_r\to e^{i\gamma_5 \theta_r/2}\psi_r$.
Then, there appears a topological term:\footnote
{Eq.\ \eqref{eq:thetaarctan} may not be applicable when $Um_r/\delta t\gg 1$ \cite{Sekine:2014xva}.}
\begin{align}
	S=\int d^4x\,\theta\frac{\alpha_e}{8\pi} F_{\mu\nu}\widetilde F^{\mu\nu},~~~~~~
	\theta\equiv \theta_0+\sum_r \theta_r=\theta_0+\sum_r \tan^{-1}\left( \frac{Um_r}{\delta t} \right),
	\label{eq:thetaarctan}
\end{align}
where $\theta_0$ is either $0$ or $1/2$ depending on the sign of $\delta t$.  (See App.\ \ref{sec:Berry} for another derivation of $\theta$.)
Note that the background magnetization $\vec m$ can fluctuate: it is a spin-wave or magnon excitation, $\vec m(\vec x)$. Then, $\theta(\vec x)$ is not a constant but a dynamical field and it has an axionic coupling to the electromagnetic field.
Therefore, in this model, the magnon effectively behaves as an axion-like field (CM axion).

\subsection{Axionic excitation as magnons}
\label{sec:cond-axion-magnon}

To relate the axionic excitation (or the CM axion) $\theta$ to the conventional magnons defined in Sec.~\ref{sec:Heisenberg}, we repeat the analysis in the previous subsection, taking into account the fluctuation of the background magnetization in terms of magnon operators.
We focus only on the spatially homogeneous spin fluctuations and consider their interaction with electrons at around a Dirac point $\vec{k} \sim \vec{k}_{\widetilde{X}_r}$.
Then, the relevant part of the Hubbard interaction term is schematically expressed as
\begin{align}
  H_U \ni U \sum_{r=1,2,3} \sum_{\vec{k} \sim \vec{k}_{\widetilde{X}_r}}
  \sum_{L=A,B} &\left[
  \widetilde{F}_L(n_{i\uparrow}; \vec{0})
  (c_{\vec{k}\downarrow,L}^\dagger c_{\vec{k}\downarrow,L}) +
  \widetilde{F}_L(n_{i\downarrow}; \vec{0})
  (c_{\vec{k}\uparrow,L}^\dagger c_{\vec{k}\uparrow,L}) \right. \notag \\
  &\left. - \widetilde{F}_L(c_{i\uparrow}^\dagger c_{i\downarrow}; \vec{0})
  (c_{\vec{k}\downarrow,L}^\dagger c_{\vec{k}\uparrow,L}) -
  \widetilde{F}_L(c_{i\downarrow}^\dagger c_{i\uparrow}; \vec{0})
  (c_{\vec{k}\uparrow,L}^\dagger c_{\vec{k}\downarrow,L}) \right],
  \label{eq:mag-Dirac-int}
\end{align}
where the Fourier transform of operators $\mathcal{O}_i$ is defined as
\begin{align}
  \widetilde{F}_L (\mathcal{O}_i; \vec{q}) \equiv
  \frac{1}{N} \sum_{i \in L} \mathcal{O}_i e^{i\vec{q} \cdot \vec{x}_i}.
\end{align}
$\tilde{F}_L$ in Eq.~\eqref{eq:mag-Dirac-int} is determined by the magnetization, which may fluctuate around the average value.
We again use the operator equations Eqs.~\eqref{eq:n_iud}--\eqref{eq:cdu} to rewrite $\tilde{F}_L$ in terms of spin operators $\vec{S}_i'$.
The relationship between $\vec{S}_i'$ and $\vec{S}_i$, which are defined in Sec.~\ref{sec:Heisenberg} and directly related to magnon operators, is given by
\begin{align}
  \vec{S}_i^{'A(B)} = O \vec{S}_i^{A(B)},
\end{align}
with $O\equiv (\vec{o}_1~ \vec{o_2}~ \vec{o}_3)$ being a $3\times 3$ rotation matrix with $\vec{m} \parallel \vec{o}_3$.\footnote{
There is an ambiguity in the choice of $\vec{o}_1$ and $\vec{o}_2$ related to the $SO(2)$ rotation around $\vec{o}_3$.
However, since \eqref{eq:magnonEB} is unchanged under the $SO(2)$ up to an overall phase factor, it does not affect the interaction strength.
}

Taking everything into consideration, the magnon-Dirac electron interaction term is, up to some constant and quadratic terms of magnons, expressed as
\begin{align}
  H_U \ni \sum_{\vec{k}} c_{\vec{k}}^\dagger \mathcal{\widetilde{H}}_U c_{\vec{k}},~~~~~
	\mathcal{\widetilde{H}}_U= \sum_{\mu=1}^5 \widetilde{R}_\mu \alpha_\mu
  + \widetilde{R}_{12} \alpha_{12}
  + \widetilde{R}_{23} \alpha_{23}
  + \widetilde{R}_{31} \alpha_{31},
  \label{eq:canonical}
\end{align}
with $\alpha_{rr'} \equiv -i \alpha_r \alpha_{r'}$.
Coefficients are given by
\begin{align}
  \widetilde{R}_r &= -U \left[ m_r + \sqrt{\frac{s}{8N}}
  \left( (O_{r1} - iO_{r2}) (u_{\vec{0}} - v_{\vec{0}})
  (\alpha_{\vec{0}} - \beta_{\vec{0}}^\dagger) + \mathrm{h.c.} \right) \right]~~(r=1,2,3),\\
  \widetilde{R}_4 &= \widetilde{R}_5 = 0,
\end{align}
where $O_{rr'}$ is the $(r,r')$ component of the rotation matrix $O$, while $m_r \equiv O_{r3} (s - \frac{1}{N} \sum_{\vec{q}} v_{\vec{q}})$ is the $r$-th component of the sublattice magnetization in the ground state.  In addition, here and hereafter, $s=1/2$.  Note that the expectation value of $\widetilde{R}_r$ is proportional to the $r$-th component of the order parameter $(\Braket{S_{i,A}} - \Braket{S_{i,B}})/2$, while that of $\widetilde{R}_{rr'}$ to the average magnetization $(\Braket{S_{i,A}} + \Braket{S_{i,B}})/2$.
The $\widetilde{R}_{rr'}$ terms induce interactions between magnon and electron/hole.  It may cause, for example, the decay of a magnon into an electron-hole pair when the gap is small.  Because we are interested in the magnon interaction with electromagnetic fields, which is not induced by the $\widetilde{R}_{rr'}$ terms, we neglect them from now on.
Repeating the same procedure as Sec.~\ref{sec:cond-axion}, we obtain the relationship between the axionic excitation and magnons.
Finally, the electromagnetic interaction of magnons is described by
\begin{align}
  H_{\text{int}} = -\frac{\alpha_e}{4\pi}\sqrt{\frac{s}{2N}}
  (u_{\vec{0}} - v_{\vec{0}})
  \left[ D^{*} \alpha_{\vec{0}}^\dagger -
  D \beta_{\vec{0}}^\dagger + \mathrm{h.c.} \right]
  \int d^3 x\, \vec{E}\cdot\vec{B},
  \label{eq:magnonEB}
\end{align}
with
\begin{align}
  D = \sum_r \frac{U/\delta t}{1 + U^2 m_r^2 / \delta t^2} (O_{r1} - iO_{r2}),
\end{align}
being an $O(1)$ factor, assuming only a moderate hierarchy between $U$ and $\delta t$.
Note that $(u_{\vec{0}} - v_{\vec{0}})$ is real because $\gamma_{\vec{0}}=1$.
The interaction Hamiltonian shows that a linear combination of magnon states is excited by a non-zero value of $\vec{E}\cdot\vec{B}$.\footnote{
    From Eq.~(\ref{eq:magnonEB}), one can read off the ``decay constant'' of the CM axion as $f_{\rm CM}\sim \left((u_{\vec 0}-v_{\vec 0})|D|\sqrt{\omega_{\vec 0} V_{\rm unit}}\right)^{-1}$ with $\omega_{\vec 0}$ being the magnon frequency at $\vec k=0$ and $V_{\rm unit}=V/N$ the volume of the magnetic unit cell.
}

\section{Dark matter conversion into condensed matter axion}
\label{sec:DM}
\setcounter{equation}{0}

Now we discuss the detection of the elementary-particle DM axion (or ALPs) and hidden photon through the interaction with CM axion.  (To avoid confusion between the DM and CM axions, hereafter, the DM axion and ALPs are both called ALPs.)

\subsection{ALP dark matter}

The dynamics of the ALP DM $a$ and the photon in a material is described by
\begin{align}
  \mathcal{L} = \frac{1}{2} (\partial_\mu a)^2
  - \frac{m_a^2}{2} a^2
  + \frac{1}{2} \left(\epsilon |\vec{E}|^2 - \frac{|\vec{B}|^2}{\mu}\right)
  + g_{a\gamma\gamma} a \vec{E} \cdot \vec{B},
\end{align}
where $\epsilon$ and $\mu$ are the permittivity and permeability of the material.
Hereafter, we treat the ALP field as a classical background
\begin{align}
  a(\vec{x}, t) = a_0 \cos (m_a t - m_a \vec{v}_a \cdot \vec{x} + \delta),
\end{align}
with $|\vec{v}_a| \sim O(10^{-3})$.
When the ALP explains the total amount of the dark matter $\rho_{\text{DM}} \sim 0.3\,\mathrm{GeV}/\mathrm{cm}^3$, we obtain $m_a^2 a_0^2 / 2 = \rho_{\text{DM}}$.
We consider applying a constant magnetic field $\vec{B}_0 = B_0 \hat{z}$ to the system, where $\hat{z}$ is a unit vector along the $z$-axis.
This magnetic field, combined with the ALP background, generates an oscillating electric field
\begin{align}
  \vec{E} (\vec{x}, t) = E_0 \hat{z} \cos(m_a t - m_a \vec{v}_a \cdot \vec{x} + \delta),
\end{align}
with
\begin{align}
  E_0 = -\frac{1}{\epsilon} g_{a\gamma\gamma} a_0 B_0.
\end{align}

The target mass range of this set up is $m_a \sim O(10^{-3})\,\mathrm{eV}$, which has a de-Broglie length $\ell \sim 1/m_a |\vec{v}_a| \sim O(10)\,\mathrm{cm}$.
We assume that $\ell$ is larger than the material size and neglect the $\vec{x}$ dependence of the ALP background inside the material.
Since $\vec{E}\cdot\vec{B}$ is uniform in this case, only the magnon zero-modes may be excited, which are considered in Sec.~\ref{sec:cond-axion-magnon}.
Substituting the value of $\vec{E}\cdot\vec{B}$ generated by the ALP background, the interaction Hamiltonian is rewritten as
\begin{align}
  H_{\mathrm{int}} = (C_a^{*} \alpha_{\vec{0}}^\dagger - C_a \beta_{\vec{0}}^\dagger + \text{h.c.})
  \cos (m_a t + \delta),
  \label{eq:Hint-axion}
\end{align}
where
\begin{align}
  C_a \equiv - \frac{\alpha_e E_0 B_0 V}{4\pi} \sqrt{\frac{s}{2N}} (u_{\vec{0}} - v_{\vec{0}}) D,
\end{align}
with $V$ being the material volume.
$H_\mathrm{int}$ describes the generation of both $\alpha$- and $\beta$-modes of the magnon.  However, as we will see below, one of them is highly enhanced when the corresponding excitation energy matches with the ALP mass; in such a case, we may expect an observable signal rate at the laboratory.
Accordingly, we will estimate a signal rate of the magnon excitation assuming that a single mode is selectively excited.\footnote{
Precisely speaking, the $\alpha$- and $\beta-$modes are not mass eigenstates since they mix with a photon, forming the so-called axionic polariton \cite{Li:2009tca}.
However, since the mixing is expected to be small for a small momentum, we neglect it in our analysis.
}

We start from the $\alpha$-mode, while the discussion for the $\beta$-mode is parallel, as we will comment later.
We define the ground and the one-magnon states of the material through $\alpha_{\vec{q}} \ket{0} = \beta_{\vec{q}} \ket{0} = 0$ for any $\vec{q}$ and $\ket{1} \equiv \alpha_{\vec{0}}^\dagger \ket{0}$, respectively.
Also, we express the state of the material at the time $t$ as\footnote
{The occupation number can be larger than $1$.  In the present case, however, the expectation value of the occupation number is much smaller than $1$, and the states with higher occupation numbers are irrelevant.}
\begin{align}
  \ket{\psi(t)} \equiv a_0(t) \ket{0} + a_1(t) \ket{1},
\end{align}
and consider its time evolution described by
\begin{align}
  i \frac{\partial}{\partial t} \ket{\psi(t)} = (H + H_{\mathrm{int}}) \ket{\psi(t)},
\end{align}
where $H$ and $H_{\mathrm{int}}$ are given in Eqs.~\eqref{eq:H0-magnon} and \eqref{eq:Hint-axion}, respectively.
We treat $H_{\mathrm{int}}$ as a perturbation and evaluate the time evolution perturbatively.
Expressing the time derivative with a dot, the evolution of coefficients $a_0(t)$ and $a_1(t)$ is described as
\begin{align}
  i \dot{a}_0 &= C_a^{*} \cos(m_a t + \delta) a_1,\\
  i \dot{a}_1 &= m_m a_1 + C_a \cos(m_a t + \delta) a_0,
\end{align}
where the magnon mass is defined as $m_m \equiv \omega_{\vec{0}} + \omega_L$.
By solving these equations, we obtain
\begin{align}
  a_1(t) \simeq -\frac{C_a}{2} \frac{ e^{i\delta}(m_a-m_m)(e^{im_at}-e^{-im_m t})
  - e^{-i\delta}(m_a+m_m)(e^{-im_at}-e^{-im_m t}) }{m_a^2-m_m^2}.
\end{align}
The probability that we find a one-magnon state $\ket{1}$ at the time $t$ is given by $P(t) \equiv |a_1(t)|^2$.
$P(t)$ is highly enhanced when $m_m \simeq m_a$, with which we obtain
\begin{align}
  P(t) \simeq \frac{|C_a|^2 t^2}{4}.
\end{align}
For the $\beta$-mode, we can repeat the discussion by defining $\ket{1} \equiv \beta_{\vec{0}}^\dagger \ket{0}$, and all the calculations are the same but replacements $C_a\to C_a^{*}$ and $\omega_L\to -\omega_L$.

$P(t)$ can not become infinitely large because there is an upper limit on $t$ for several reasons; one of them is the ALP coherence time $\tau_a \sim 1/m_a v_a^2$ and another is the magnon dissipation time $\tau_m$.
Neglecting other possible sources of limitation for simplicity, we define the effective coherence time $\tau \equiv \min (\tau_a, \tau_m)$.
Then, the average magnon excitation rate is evaluated as
\begin{align}
  \frac{d N_{\mathrm{signal}}}{d t} = \frac{P(\tau)}{\tau} = \frac{|C_a|^2 \tau}{4}.
  \label{eq:signal-rate}
\end{align}
Numerically, the signal rate is evaluated as
\begin{align}
  \frac{d N_{\mathrm{signal}}}{d t} \sim 0.002\,\mathrm{s}^{-1}
  &\left( \frac{B_0}{1\,\mathrm{T}} \right)^4
  \left( u_{\vec{0}} - v_{\vec{0}} \right)^2
  \left( \frac{V_{\text{unit}}}{(0.3\,\mathrm{keV})^{-3}} \right)
  \left( \frac{V}{(10\,\mathrm{cm})^3} \right)
  \notag \\
  \times
  &\frac{|D|^2}{\epsilon^2}
  \left( \frac{g_{a\gamma\gamma}}{10^{-10}\,\mathrm{GeV}^{-1}} \right)^2
  \left( \frac{10^{-3}\,\mathrm{eV}}{m_a} \right)^2
  \left( \frac{\tau}{0.1\,\mathrm{\mu s}} \right),
  \label{eq:R-axion}
\end{align}
where $V/N = V_{\mathrm{unit}}$ with $V_{\mathrm{unit}}$ being the volume of the magnetic unit cell.
Note that, from Eq.~(\ref{eq:u2v2}), a straightforward calculation shows
\begin{align}
  (u_{\vec{0}} - v_{\vec{0}})^2 =
  \sqrt{\frac{2\omega_J + \omega_A}{\omega_A}},
\end{align}
and hence the signal rate is enhanced if $\omega_J \gg \omega_A$.

\begin{figure}[t]
  \centering
  \includegraphics[width=0.6\hsize]{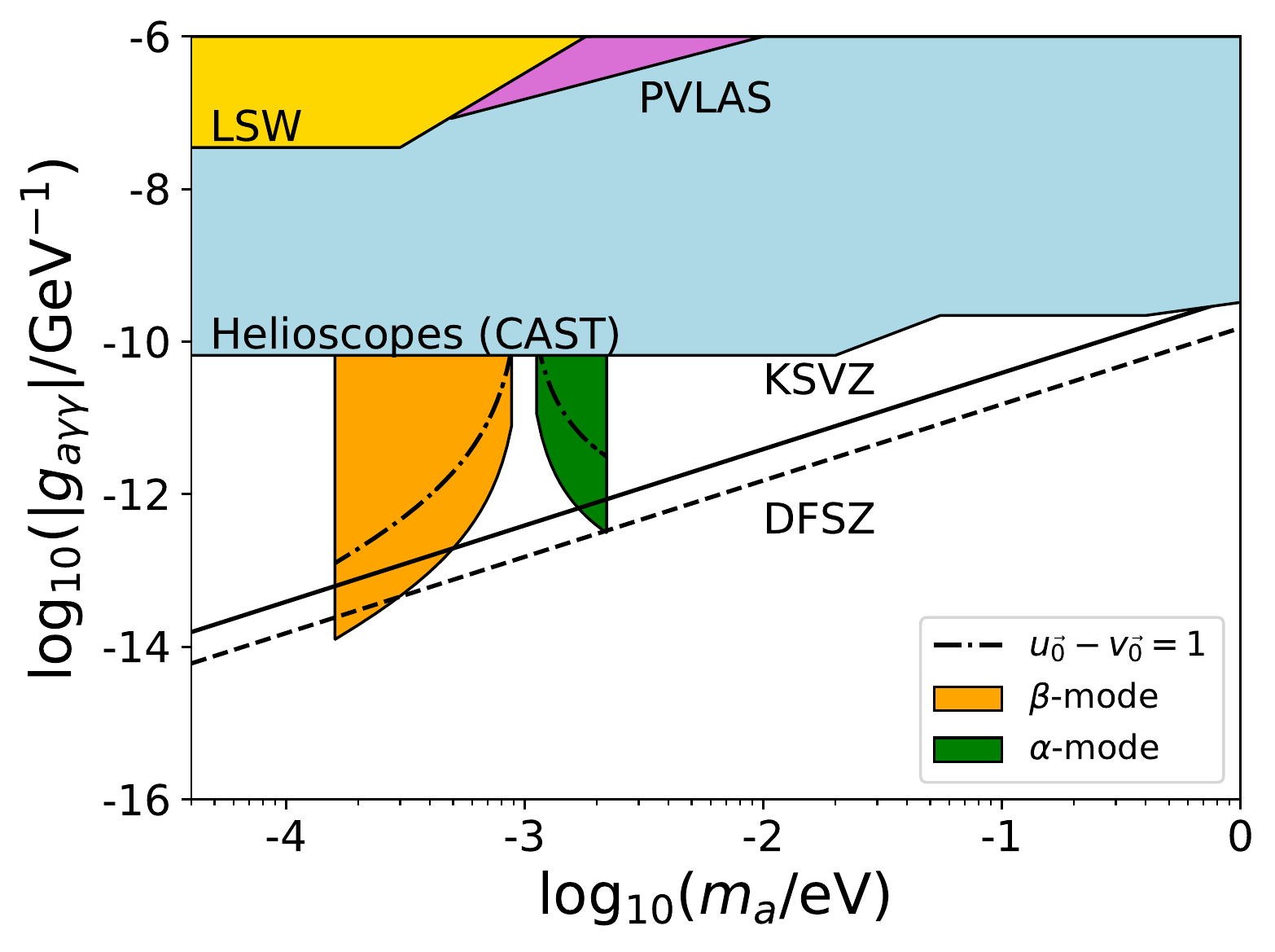}
  \caption{
  Sensitivity of the magnon to the ALP DM in the $m_a$ vs. $g_{a\gamma\gamma}$ plane.
  The orange (green) region corresponds to the sensitivity of the $\beta$-mode ($\alpha$-mode) with $u_{\vec{0}}-v_{\vec{0}}=10$, while the dot-dashed line in each region shows the sensitivity of the corresponding mode with $u_{\vec{0}}-v_{\vec{0}}=1$.
  We postulate the target volume $V=(10\,\mathrm{cm})^3$ and the magnetic field scanned over $1\,\mathrm{T} < B_0 < 7\,\mathrm{T}$ ($1\,\mathrm{T} < B_0 < 10\,\mathrm{T}$) for the $\beta$-mode ($\alpha$-mode).
  For each step of the scan, we use $\Delta t = 10^2\,\mathrm{s}$ for an observation, which requires $\sim 1\,\mathrm{yr}$ for the whole scan.
  See the text for more details of the material properties.
  Also shown as colored regions are existing constraints, while the black solid (dashed) line shows the prediction for the KSVZ (DFSZ) model.
  }
  \label{fig:axion}
\end{figure}

In Fig.~\ref{fig:axion}, we show the sensitivity on the ALP parameter space taking $(u_{\vec{0}}-v_{\vec{0}}) = 1$ and $10$, $V_{\text{unit}} = (0.3\,\mathrm{keV})^{-3}$, and $|D|^2=\epsilon=1$ as the material properties and postulating $V = (10\,\mathrm{cm})^3$.
We also assume $\tau_a < \tau_m$ and use
\begin{align}
  \tau = \frac{1}{m_a v_a^2} \sim 0.7\,\mathrm{\mu s} \left( \frac{10^{-3}\,\mathrm{eV}}{m_a} \right).
\end{align}
As for the magnon dispersion relation, we use typical values
\begin{align}
  m_m = 1.0 \pm 0.12 \left( \frac{B_0}{1\,\mathrm{T}} \right)\,\mathrm{meV},
\end{align}
where the plus (minus) sign is selected for the $\alpha$- ($\beta$-)mode.
The magnetic field is assumed to be scanned within the range $1\,\mathrm{T} < B_0 < 10\,\mathrm{T}$.
The $\beta$-mode is used for our analysis only when $B_0 < 7\,\mathrm{T}$ to avoid the instability or the enhanced noise rate according to the low frequency.
For each step of the scan, we can search for a mass range of $\Delta m_a \sim 2/\tau \sim 10^{-8}\,\mathrm{eV}$ and we use $\Delta t_{\text{scan}} \sim 10^2\,\mathrm{s}$ for an observation.
Accordingly, in order to cover all the accessible ALP mass, it takes $\sim 1\,\mathrm{year}$ to scan the magnetic field.
We do not discuss in detail the detection method of generated magnons in this paper; they might be observed through the conversion into photons at the boundary of the material as in \cite{Marsh:2018dlj}, or might be detected using some specific features for axionic insulators, such as the dynamical chiral magnetic effect \cite{Sekine:2020ixs}.
For the estimation of the sensitivity, we just assume the noise rate for the detection $dN_{\text{noise}}/dt \sim 10^{-3}\,\mathrm{s}^{-1}$ as is adopted in \cite{Marsh:2018dlj}, which is an already demonstrated value for a single photon detector in the $\mathrm{THz}$ regime at the temperature $T=0.05\,\mathrm{K}$ \cite{Komiyama:2000}.
We estimate the sensitivity by requiring the signal-to-noise ratio (SNR)
\begin{align}
        (\text{SNR}) \equiv \frac{(d N_{\text{signal}}/dt)\, \Delta t_{\text{scan}}}{\sqrt{(d N_{\text{noise}}/dt)\, \Delta t_{\text{scan}}}},
\end{align}
to be larger than $3$ for each scan step.

In the figure, the orange and green regions correspond to the sensitivity using $\beta$- and $\alpha$-modes, respectively, with $u_{\vec{0}}-v_{\vec{0}}=10$, while the dot-dashed line in each region shows the sensitivity of the corresponding mode with $u_{\vec{0}}-v_{\vec{0}}=1$.
The other colored regions show existing constraints from the Light-Shining-through-Walls (LSW) experiments such as the OSQAR~\cite{Ballou:2015cka} (yellow), the measurement of the vacuum magnetic birefringence at the PVLAS~\cite{DellaValle:2015xxa} (pink), and the observation of the ALP flux from the sun using the helioscope CAST~\cite{Anastassopoulos:2017ftl} (blue).
We also show the predictions of the KSVZ and DFSZ axion models with black solid and dashed lines, respectively.
We can see that the use of both $\alpha$- and $\beta$-modes gives a detectability over a broad mass range of $10^{-3}$--$10^{-2}\,\mathrm{eV}$ and the sensitivity may reach both the KSVZ and DFSZ model predictions for some mass range.
It is also notable that the sensitivity becomes much better for the lighter (heavier) mass region with the $\beta$-mode ($\alpha$-mode), both of which correspond to larger $B_0$, due to the $B_0^4$ dependence of the signal rate.

\subsection{Hidden photon dark matter}

We consider a hidden $U(1)$ gauge field $H_\mu$, which has a kinetic mixing with the $U(1)_Y$ hypercharge gauge boson $B_\mu$.
The relevant Lagrangian is
\begin{align}
  \mathcal{L} = -\frac{1}{4} H_{\mu\nu} H^{\mu\nu}
  -\frac{1}{4} B_{\mu\nu} B^{\mu\nu}
  +\frac{\epsilon_Y}{2} H_{\mu\nu} B^{\mu\nu}
  +\frac{1}{2} m_H^2 H_\mu H^\mu,
\end{align}
where $m_H$ is the hidden photon mass.
Below, we use the convention that the expressions such as $H_{\mu\nu}$ and $B_{\mu\nu}$ denote the field strengths of the corresponding gauge fields $H_\mu$ and $B_\mu$, respectively.
After redefining fields as $B'_\mu \equiv B_\mu - \epsilon_Y H_\mu$ and $H'_\mu \equiv \sqrt{1-\epsilon_Y^2} H_\mu$, we can rewrite the kinetic terms in the canonical form and obtain
\begin{align}
  \mathcal{L} = -\frac{1}{4} H'_{\mu\nu} H'^{\mu\nu}
  -\frac{1}{4} B'_{\mu\nu} B'^{\mu\nu}
  +\frac{1}{2} m_{H'}^2 H_\mu' H'^\mu,
\end{align}
with $m_{H'} \equiv m_H / \sqrt{1-\epsilon_Y^2}$.
After the electroweak symmetry breaking, there appear additional mass terms and further mixing occurs.
The mass terms are given by
\begin{align}
  \mathcal{L}_{\text{mass}} = \frac{m_Z^2}{2} (c_W W_\mu^3 - s_W B_\mu)^2
  + \frac{m_{H'}^2}{2} H'_\mu H'^\mu,
\end{align}
where $m_Z$ is the $Z$-boson mass, $W_\mu^3$ is the third component of the $SU(2)_L$ gauge bosons, while $c_W \equiv \cos \theta_W$ and $s_W \equiv \sin \theta_W$ with $\theta_W$ being the Weinberg angle.
The mass terms are approximately diagonalized by performing the unitary transformation
\begin{align}
  \begin{pmatrix}
    W_\mu^3 \\
    B'_\mu \\
    H'_\mu
  \end{pmatrix} = \begin{pmatrix}
    c_W & -s_W & s_W c_W \epsilon_Y \\
    -s_W & c_W & s_W^2 \epsilon_Y \\
    -s_W \epsilon_Y & 0 & 1
  \end{pmatrix} \begin{pmatrix}
    Z_\mu \\
    A_\mu \\
    H''_\mu
  \end{pmatrix},
\end{align}
up to terms of $O(\epsilon_Y m_H'^2)$ and $O(\epsilon_Y^2 m_Z^2)$.
The mass-squared eigenvalues are $m_Z^2$, $0$, and $m_H'^2$ for $Z_\mu$, $A_\mu$, and $H''_\mu$ fields, respectively.

According to the mixing among gauge bosons described above, the interaction between $H''_\mu$ and electrons is induced as
\begin{align}
  \mathcal{L}_{\text{int}} = - \epsilon_H e H''_\mu \bar{\psi} \gamma^\mu \psi,
\end{align}
where $\epsilon_H \equiv \epsilon_Y c_W$ and $\psi$ is an electron field.
Since the electromagnetic interaction of magnons \eqref{eq:magnonEB} originates from the triangle diagram of Dirac electrons, a hidden photon field can replace a photon field in the interaction at the cost of a factor $\epsilon_H$, leading to the magnon-hidden photon-photon interaction
\begin{align}
  H_{\text{int}} = -\frac{\epsilon_H \alpha_e}{4\pi}\sqrt{\frac{s}{2N}}
  (u_{\vec{0}} - v_{\vec{0}})
  \left[ D^{*} \alpha_{\vec{0}}^\dagger -
  D \beta_{\vec{0}}^\dagger + \mathrm{h.c.} \right]
  \int d^3 x\, \vec{E}_H\cdot\vec{B},
\end{align}
with $\vec{E}_H \equiv -\vec{\nabla} H''_0 - \dot{\vec{H}}''$ being the hidden electric field.

From now on, let us resort to the abbreviation of $H_\mu$ and $m_H$ for the mass eigenstate and eigenvalue of the hidden photon for notational simplicity.
We consider the light hidden photon to explain the whole amount of the DM.\footnote{
	The correct relic abundance of hidden photon DM of ${\rm meV}$ mass range is reasonably explained by the gravitational production mechanism~\cite{Graham:2015rva,Ema:2019yrd,Ahmed:2020fhc,Kolb:2020fwh} or the production from cosmic strings~\cite{Long:2019lwl}.
}
Taking into account the equation of motion $(\Box + m_H^2) H_\mu = 0$ and $\partial_\mu H^\mu = 0$, we can express each component of the hidden photon field as
\begin{align}
  H_0 (t, \vec{x}) &= -\vec{v}_H \cdot \vec{\tilde{H}} \cos(m_H t - m_H \vec{v} \cdot \vec{x} + \delta),\\
  \vec{H} (t, \vec{x}) &= \vec{\tilde{H}} \cos(m_H t - m_H \vec{v} \cdot \vec{x} + \delta),
\end{align}
with $\rho_{\text{DM}} = m_H^2 \tilde{H}^2 / 2$ and $\tilde{H} \equiv |\vec{\tilde{H}}|$.
In this parametrization, the hidden electric field is expressed as
\begin{align}
  \vec{E}_H = \vec{\tilde{H}} m_H \sin (m_H t + \delta).
\end{align}

By repeating the same analysis as in the previous subsection, we can estimate the magnon excitation rate from the existence of the hidden photon coherent oscillation.
The rate is given by $dN_{\text{signal}}/dt = |C_H|^2 \tau / 4$ with
\begin{align}
  C_H = - \frac{\alpha_e \tilde{H} m_H B_0 V}{4\pi} \cos\theta \sqrt{\frac{s}{2N}} (u_{\vec{0}} - v_{\vec{0}}) D,
\end{align}
and $\tau = \min(\tau_H, \tau_m)$ with $\tau_H \sim 1/m_H v_H^2$.
$\theta$ is defined as an angle between $\vec{\tilde{H}}$ and $\vec{B}_0$.
Numerically, we obtain the estimation
\begin{align}
  \frac{d N_{\mathrm{signal}}}{d t} \sim 0.02\,\mathrm{s}^{-1}
  &\left( \frac{B_0}{1\,\mathrm{T}} \right)^2
  \left( u_{\vec{0}} - v_{\vec{0}} \right)^2
  \left( \frac{V_{\text{unit}}}{(0.3\,\mathrm{keV})^{-3}} \right)
  \left( \frac{V}{(10\,\mathrm{cm})^3} \right)
  \notag \\
  \times
  &|D|^2
  \left( \frac{\epsilon_H}{10^{-13}} \right)^2
  \left( \frac{\cos^2\theta}{1/2} \right)
  \left( \frac{\tau}{0.1\,\mathrm{\mu s}} \right).
\end{align}
Note that the signal rate is proportional to a different power of the magnetic field and the DM mass compared with that for the ALP \eqref{eq:R-axion}.

\begin{figure}
  \centering
  \includegraphics[width=0.6\hsize]{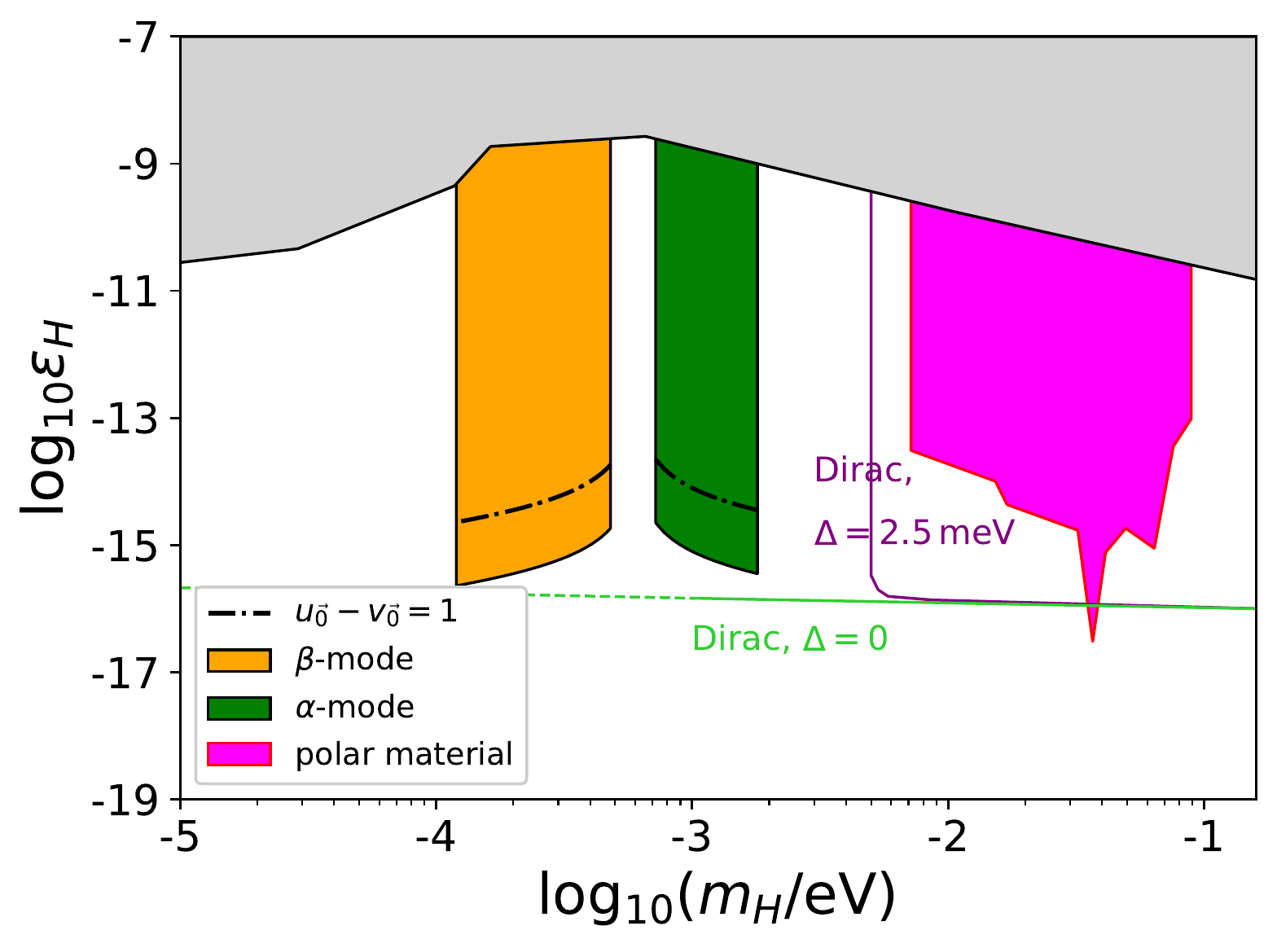}
  \caption{
  Sensitivity of the magnon to the DM hidden photon in the $m_H$ vs. $\epsilon$ plane.
  The color and line style convention and the experimental set up are the same as those explained in Fig.~\ref{fig:axion}.
  The gray region is a combination of existing constraints, while the magenta region shows a sensitivity of the polar material~\cite{Knapen:2017ekk}.
  The purple and green lines correspond to the sensitivity of the Dirac material~\cite{Hochberg:2017wce} with gap sizes $\Delta = 2.5\,\mathrm{meV}$ and $0$, respectively.
  }
  \label{fig:hidden-photon}
\end{figure}

In Fig.~\ref{fig:hidden-photon}, we show the sensitivity in the hidden photon parameter space.
The assumptions for the material properties are the same as those used in the previous subsection, while we assume $\tau = \tau_H \sim 1/m_H v_H^2$ and $\cos^2 \theta = 1/2$ in this case.
Again, the orange and green regions correspond to the sensitivity of $\beta$- and $\alpha$-modes, respectively.
The gray region shows existing constraints taken from~\cite{McDermott:2019lch}, while the magenta region shows a sensitivity of the proposal with a polar material~\cite{Knapen:2017ekk}.
The purple and green lines correspond to the sensitivity of the Dirac material~\cite{Hochberg:2017wce} with gap sizes $\Delta = 2.5\,\mathrm{meV}$ and $0$, respectively.
We can see that the use of magnons gives a good sensitivity over a mass range $10^{-3}$--$10^{-2}\,\mathrm{eV}$ of the hidden photon.
The sensitivity has a smaller mass dependence compared with the result for the ALP because of the smaller power of $B_0$ in the expression of the signal rate.

\section{Conclusions and discussion}
\label{sec:conc}
\setcounter{equation}{0}

Motivated by recent developments in the axion electrodynamics in the context of condensed matter physics, we considered a possibility of DM detection through DM conversion into the condensed-matter (CM) axion. 
We formulated a way how the CM axion degree of freedom appears starting from the tight-binding model of the electrons on the lattice. In a particular example, we have taken the model in~\cite{Sekine:2014xva}, in which the CM axion may be interpreted as the spin wave or the (linear combination of) magnons in an anti-ferromagnetic insulator.\footnote{
	In the original proposal of dynamical axion in Fe-doped topological insulators such as ${\rm Bi_2 Se_3}$~\cite{Li:2009tca}, the CM axion is interpreted as an amplitude mode of the anti-ferromagnetic order parameter and not expressed by a linear combination of magnons.
} 
For the convenience of readers of particle physics side, we have reviewed the Heisenberg model and half-filling Hubbard model in a self-consistent and comprehensive manner. Based on these basic ingredients, we can derive the CM axion dispersion relation and its interaction with electromagnetic fields. 

As DM models, we considered two cases: the elementary particle axion (or ALP) and the hidden photon. We calculated the DM conversion rate into the CM axion in a quantum mechanical way and estimated the signal rate. 
It is possible to cover the parameter regions which have not been explored so far in the DM mass range of about meV.  It may be possible to reach the QCD axion.
One should note, however, that our calculation is just based on an idealized theoretical model of the electron system in the anti-ferromagnetic insulator. It is nontrivial how well such a description is when it is applied to a real material. 
We have not provided a concrete way to detect the CM axion excitation. One possible way is to use the photon emission through the CM axion-photon mixing (axionic polariton) and detect it by the dish antenna as discussed in Ref.~\cite{Marsh:2018dlj}.
In any case, it is important to understand the origin of CM axion and its properties, and we believe our formulation gives a basis of the estimation of the CM axion production rate from background DM and is useful for future developments of this field.

A final comment is that the physics of CM axion is very rich and the CM axion in a different material may have a different microphysical origin \cite{Sekine:2020ixs,Nenno:2020ujq}. It would be interesting to explore the physics of CM axion as a probe of DM in a broader class of materials.

\section*{Note added}

While finalizing this manuscript, a related paper appeared on arXiv~\cite{1845816}.

\section*{Acknowledgments}
This work was supported by JSPS KAKENHI Grant (Nos. 20J00046 [SC], 16H06490 [TM], 18K03608 [TM], 18K03609 [KN] and 17H06359 [KN]).
SC was supported by the Director, Office of Science, Office of High Energy Physics of the U.S. Department of Energy under the Contract No. DE-AC02-05CH1123.

\appendix
\section{Note on spin-orbit interaction term}
\label{sec:spin}
\setcounter{equation}{0}

In this appendix, we see how to derive the spin-orbit interaction term given in Eq.~\eqref{eq:TB-SO}.
We will first discuss how the hamiltonian is expressed in terms of creation and annihilation operators of the electron.
Next, we derive the effective hamiltonian of graphene as an example, which becomes the same form as \eqref{eq:TB-SO}, and then show that the result is model independent.

\subsection{Tight-binding model with spin-orbit interaction}

\begin{table}[t]
  \centering
  \begin{tabular}{c|cccc}
    $\mu \setminus \nu$ & $s$ & $p_x$ & $p_z$ & $d_{zx}$ \\ \hline
    $s$ & $V_{ss\sigma}$ & $n_x V_{sp\sigma}$ & $n_z V_{sp\sigma}$ & $\sqrt{3}n_xn_z V_{sd\sigma}$ \\
    $p_x$ & $*$ & $n_x^2 V_{pp\sigma} + (1-n_x^2) V_{pp\pi}$ & $n_xn_z V_{pp\sigma} - n_xn_z V_{pp\pi}$ & $\sqrt{3}n_x^2n_z V_{pd\sigma} + n_z(1-2n_x^2) V_{pd\pi}$ \\
    $p_z$ & $*$ & $*$ & $n_z^2 V_{pp\sigma} + (1-n_z^2)V_{pp\pi}$ & $n_x V_{pd\pi}$ \\
    $d_{zx}$ & $*$ & $*$ & $*$ & $n_x^2 V_{dd\pi} + n_y^2 V_{dd\delta}$
  \end{tabular}
  \caption{
  Table of off-diagonal elements of $T_{\mu\nu}^{ij}$ \cite{PhysRev.94.1498}.
  $\vec{n} \equiv \vec{r}_j - \vec{r}_i$ denotes the lattice displacement vector.
  We omitted the principal quantum numbers associated with $\mu$ and $\nu$ since a different choice only results in different numerical values of $V$-factors such as $V_{ss\sigma}$.
  The left bottom elements with $*$ markers can be obtained by the relationship $T_{\mu\nu}^{ij} = T_{\nu\mu}^{ji} = \left. T_{\nu\mu}^{ij} \right|_{\vec{n}\to-\vec{n}}$.
  }
  \label{tab:Slater-Koster}
\end{table}

We consider a model in which atoms are attached to lattice points labeled by $i$ with position vectors $\vec{r}_i$.
Each atom has its energy eigenstates generated by $c_{\mu i}^\dagger$, where $\mu$ denotes an electron orbital.
The diagonal part of the tight-binding hamiltonian, $H_{\mathrm{TB}}$, is given by the sum of the hamiltonian of each atom.
On the other hand, a small overlap between electron wave functions sit at different lattice sites induces relatively small off-diagonal elements.
We are particularly interested in the case where electrons in each atom are tightly bound on a lattice point.
In this case, we can neglect the overlap between two sites unless they are the nearest neighbors of each other.
Accordingly, we obtain
\begin{align}
  H_{\mathrm{TB}} =
  \sum_{\mu} \sum_i \epsilon_\mu c_{\mu i}^\dagger c_{\mu i} +
  \sum_{\mu,\nu} \sum_{\Braket{i,j}} T_{\mu\nu}^{ij} c_{\mu i}^\dagger c_{\nu j},
  \label{eq:TB}
\end{align}
where $\epsilon_\mu$ denotes the energy level of the electron orbital $\mu$ of a single atom.\footnote{
In general, the energy level may change against the choice of the atom.
However, we only focus on the case where it is universal for all the atoms in this paper.
}
The off-diagonal elements $T_{\mu\nu}^{ij}$ are calculated by Slater and Koster \cite{PhysRev.94.1498} as summarized in Table~\ref{tab:Slater-Koster} for several important choices of electron orbitals.
One of the important features of these results is the directional dependence (i.e., the existence of $\vec{n}\equiv \vec{r}_j - \vec{r}_i$ in the expressions), which is sourced from the directional dependence of orbitals.
Information of the shape of the lattice comes into the Hamiltonian due to this dependence.

\begin{table}[t]
  \centering
  \begin{tabular}{cc}
    \begin{tabular}[t]{c|cccc}
      $\mu \setminus \nu$ & $s$ & $p_x$ & $p_y$ & $p_z$ \\ \hline
      $s$ & $0$ & $0$ & $0$ & $0$ \\
      $p_x$ & $0$ & $0$ & $-is_z$ & $is_y$ \\
      $p_y$ & $0$ & $is_z$ & $0$ & $-is_x$ \\
      $p_z$ & $0$ & $-is_y$ & $is_x$ & $0$
    \end{tabular}
    &
    \begin{tabular}[t]{c|ccccc}
      $\mu \setminus \nu$ & $d_{xy}$ & $d_{x^2-y^2}$ & $d_{zx}$ & $d_{yz}$ & $d_{z^2}$ \\ \hline
      $d_{xy}$ & 0 & $2is_z$ & $-is_x$ & $is_y$ & $0$ \\
      $d_{x^2-y^2}$ & $-2is_z$ & $0$ & $is_y$ & $is_x$ & $0$ \\
      $d_{zx}$ & $is_x$ & $-is_y$ & $0$ & $-is_z$ & $i\sqrt{3}s_y$ \\
      $d_{yz}$ & $-is_y$ & $-is_x$ & $is_z$ & $0$ & $-i\sqrt{3}s_x$ \\
      $d_{z^2}$ & $0$ & $0$ & $-i\sqrt{3}s_y$ & $i\sqrt{3}s_x$ & $0$ \\
    \end{tabular}
  \end{tabular}
  \caption{
  Summary of the matrix elements $\Braket{\vec{L}\cdot \vec{S}}_{\mu\nu}$.
  It is implicitly assumed that the principal quantum numbers of $\mu$ and $\nu$ are the same.
  The left (right) panel shows the results for $s$ and $p$ ($d$) orbitals.
  Note that the spin operators are related to the Pauli matrices as $s_f = \sigma_f/2$ ($f=x,y,z$).
  }
  \label{tab:LS}
\end{table}

Next, we take into account the effects of the spin-orbit interaction.
Due to the relativistic motion of an electron inside an atom, it feels a magnetic field whose size and direction are proportional to its angular momentum $\vec{L}$.
As a result, we obtain the on-site spin-orbit interaction Hamiltonian
\begin{align}
  H_{\mathrm{SO}} = \frac{1}{m_e^2 r} \frac{d V(r)}{d r} \vec{L}\cdot \vec{S},
  \label{eq:HSO}
\end{align}
where $V(r)$ is the centrifugal potential in which the electron moves, while $\vec{L}$ and $\vec{S}$ are the electron angular momentum and spin operators, respectively.
Given that the operator $\vec{L}\cdot \vec{S}$ does not change the principal and azimuthal quantum numbers, this interaction induces the term
\begin{align}
  H_{\mathrm{SO}} = \sum_{\mu,\nu} \sum_i \xi_{n\ell} c_{\mu i}^\dagger \Braket{\vec{L}\cdot \vec{S}}_{\mu\nu} c_{\nu i},
  \label{eq:H_SO}
\end{align}
where $n$ and $\ell$ are the common principal and azimuthal quantum numbers of $\mu$ and $\nu$, respectively, while $\xi_{n\ell}$ denotes the radial average of the coefficient in Eq.~\eqref{eq:HSO}.
Some of the matrix elements $\Braket{\vec{L}\cdot \vec{S}}_{\mu\nu}$ are shown in Table~\ref{tab:LS} as examples.

\subsection{Graphene}

Graphene is made of carbon atoms that are located on the two-dimensional honeycomb lattice on the $xy$ plane.
Three out of four electrons of the outermost shell of each carbon in $2s$, $2p_x$, and $2p_y$ orbitals are shared among the nearest neighbor carbons to form the so-called $\sigma$ bond.
On the other hand, the other electron in the $2p_z$ orbital is also shared and called the $\pi$ bond.
The unit cell consists of two lattice sites, which we call $A$ and $B$ sublattices.
Since we are particularly interested in the dynamics of electrons in $p_z$ orbitals of $A$ and $B$ sublattices, we construct an effective theory of electron states in $p_z$ orbitals by integrating out all the other states.

Among the full hamiltonian $H \equiv H_{\mathrm{TB}} + H_{\mathrm{SO}}$, we treat the off-diagonal elements, i.e., the second term of Eq.~\eqref{eq:TB} and $H_{\mathrm{SO}}$, as perturbations and name the corresponding part of $H$ as $V$.
Also, we call an effective theory hamiltonian $H_{\mathrm{eff}}$ and its off-diagonal part $V_{\mathrm{eff}}$, both of which are constructed only from $c_{p_z, i}$ and $c_{p_z, i}^\dagger$.
Then, the matching condition of the full theory to the effective theory is given by
\begin{align}
  \Braket{2p_z, i | U^\dagger (t, t_0) | 2p_z, j} =
  \Braket{2p_z, i | U^\dagger_{\mathrm{eff}} (t, t_0) | 2p_z, j},
  \label{eq:matching}
\end{align}
where $\ket{\mu, i} \equiv c_{\mu i}^\dagger \ket{0}$ with $\ket{0}$ being the vacuum state, while $U$ and $U_{\mathrm{eff}}$ are the time evolution operators in the full and effective theories, respectively.
Working in the interaction picture, they are given by
\begin{align}
  U(t,t_0) = T \left\{ \exp \left[
  i \int_{t_0}^{t} dt'\, V_I (t')
  \right] \right\},
\end{align}
with $T$ being the time-ordering operator, and
\begin{align}
  V_I(t) \equiv e^{i H_0 t} V e^{-i H_0 t},
\end{align}
while $U_{\mathrm{eff}}$ can be obtained by substituting $V$ with $V_{\mathrm{eff}}$.

The left-handed side of Eq.~\eqref{eq:matching} does not have a contribution from $H_{\mathrm{SO}}$ at the first order of $V_{I}$ since $\vec{L}\cdot \vec{S}$ does not have a non-zero matrix element.
Also, there are contributions only with even numbers of $\vec{L}\cdot \vec{S}$ at the second order of perturbation.
Such contributions just slightly modify $\epsilon_\mu$ and $T_{\mu\nu}^{ij}$ and do not qualitatively change the physics, so we just neglect it.
The third order contribution can be rewritten as
\begin{align}
  \int_{t_0}^t dt'\, \int_{t'}^t dt''\, \int_{t''}^t dt'''\,
  \sum_{\mu,\nu,k,p}
  \Braket{2p_z, i | i V_I (t') | \mu, k}
  \Braket{\mu, k | i V_I (t'') | \nu, p}
  \Braket{\nu, p | i V_I (t''') | 2p_z, j}.
  \label{eq:SOterms}
\end{align}
According to \cite{PhysRevB.82.245412}, it is known that the contributions from the spin-orbit interaction among $3d$ orbitals are numerically large in this model, so we may focus only on them.
As a result, we deform \eqref{eq:SOterms} to obtain
\begin{align}
  \left. \Braket{2p_z, i | U^\dagger (t, t_0) | 2p_z, j} \right|_{\text{3rd order in }V_I}
  \simeq -(t-t_0) \frac{\xi_{3d} V_{pd\pi}^2}{(\epsilon_{3d} - \epsilon_{2p})^2} \vec{s} \cdot (\vec{d}_{ij}^1 \times \vec{d}_{ij}^2),
\end{align}
where $\vec{d}_{ij}^1 \equiv \vec{r}_k - \vec{r}_i$ and $\vec{d}_{ij}^2 \equiv \vec{r}_j - \vec{r}_k$.
The factor $\vec{d}_{ij}^1 \times \vec{d}_{ij}^2$ forces the matrix element to be zero when $i=j$ and the only non-zero matrix elements are those with $(i,j)$ being a pair of next-nearest neighbors.
Therefore, the subscript $k$ in the definition of $\vec{d}_{1,2}$ should be understood as the lattice site in between $i$ and $j$.
The corresponding matrix element in the right-handed side of Eq.~\eqref{eq:matching} is given by
\begin{align}
  \Braket{2p_z, i | U^\dagger_{\mathrm{eff}} (t, t_0) | 2p_z, j} \simeq
  i (t-t_0) \Braket{p_z, i | V_{\mathrm{eff}} | p_z, j},
\end{align}
so we conclude
\begin{align}
  V_{\mathrm{eff}} \ni i \frac{\xi_{3d} V_{pd\pi}^2}{(\epsilon_{3d} - \epsilon_{2p})^2}
  \sum_{\Braket{\Braket{i, j}}} c_{2p_z, i}^\dagger\, \vec{s} \cdot (\vec{d}_{ij}^1 \times \vec{d}_{ij}^2) c_{2p_z, j}.
\end{align}
This agrees with \eqref{eq:TB-SO} when we set $\lambda = a^2 \xi_{3d} V_{pd\pi}^2 / 4(\epsilon_{3d} - \epsilon_{2p})^2$.

\subsection{Model independence of the spin-orbit interaction term}

So far, we have considered the spin-orbit interaction for a specific
choice of the lattice structure, i.e., the two-dimensional honeycomb
lattice.  Here, we argue that the structure of the interaction, given
in Eq.~\eqref{eq:TB-SO}, can be understood by symmetries.

\begin{figure}[t]
  \begin{center}
    \includegraphics[width=0.3\textwidth]{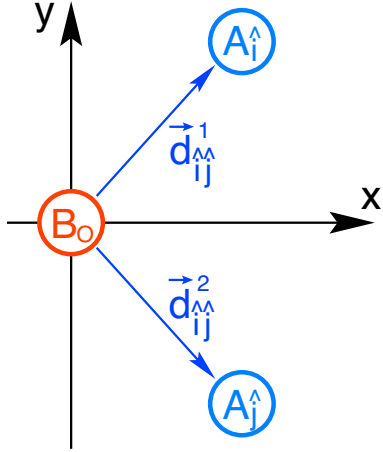}
    \caption{The coordinate adopted in deriving the general form of
      the spin-orbit interaction.}
    \label{fig:spinorbit}
    \end{center}
\end{figure}

Here, we consider the interaction between next-nearest neighbor sites
induced by the spin-orbit interaction $H_{\mathrm{SO}}$.  For this
purpose, we consider a set of next-nearest neighbor sites from the
$A$-sublattice (called $A_{\hat{i}}$ and $A_{\hat{j}}$), which share
only one nearest neighbor site (called $B_O$).  The vectors pointing
to $A_{\hat{i}}$ and $A_{\hat{j}}$ from $B_O$ are denoted as
$\vec{d}_{\hat{i}\hat{j}}^1$ and $\vec{d}_{\hat{i}\hat{j}}^2$,
respectively.  Here, we adopt a coordinate in which $A_{\hat{i}}$, $A_{\hat{j}}$, and
$B_O$ are on the $x$ vs.\ $y$ plane; the position of $B_O$ is set to
be the origin and the $y$ axis is chosen to be parallel to
$\vec{d}_{\hat{i}\hat{j}}^1-\vec{d}_{\hat{i}\hat{j}}^2$ (see
Fig.\ \ref{fig:spinorbit}).

Hereafter, we assume that the whole lattice is invariant under the
following transformations and hence the Hamiltonian also is:
\begin{itemize}
\item $\mathcal{P}$: Parity, defined as the reflection with respect to
  the $x$ vs.\ $y$ plane: $(x,y,z)\xrightarrow{\mathcal{P}} (x,y,-z)$.
  With the $\mathcal{P}$ transformation, the angular momentum operator
  acting on the electron on $i$-th site transforms as
  $(L_x^{(i)},L_y^{(i)},L_z^{(i)})\xrightarrow{\mathcal{P}}
  (-L_x^{(\mathcal{P}[i])},-L_y^{(\mathcal{P}[i])},L_z^{(\mathcal{P}[i])})$,
  where $i\xrightarrow{\mathcal{P}}\mathcal{P}[i]$.  (Thus,
  $\mathcal{P}[\hat{i}]=\hat{i}$.)  In addition, the annihilation
  operator of the electron transforms as
  \begin{align}
    c_{\mu,i} \xrightarrow{\mathcal{P}}
    \sigma_3 c_{\mathcal{P}[\mu],\mathcal{P}[i]},
  \end{align}
  where $\mathcal{P}[\mu]$ denotes the $\mathcal{P}$-transformed orbital
  of $\mu$.  (If $\mu$ is singlet under the $\mathcal{P}$-transformation,
  $\mathcal{P}[\mu]=\mu$.)
\item $\mathcal{R}$: $\pi$ rotation around the $x$ axis:
  $(x,y,z)\xrightarrow{\mathcal{R}} (x,-y,-z)$.  With this
  transformation, the lattice site $i$ is moved to the position of
  $\mathcal{R}[i]$.  With $\mathcal{R}$, the angular momentum operator
  transforms as
  $(L_x^{(i)},L_y^{(i)},L_z^{(i)})\xrightarrow{\mathcal{R}}
  (L_x^{\mathcal{R}[j]},-L_y^{\mathcal{R}[j]},-L_z^{\mathcal{R}[j]})$.
  In addition,
  \begin{align}
    c_{\mu,i} \xrightarrow{\mathcal{R}}
    \sigma_1 c_{\mathcal{R}[\mu],\mathcal{R}[i]},
  \end{align}
  where $\mathcal{R}[\mu]$ denotes the $\mathcal{R}$-transformed
  orbital of $\mu$.
\end{itemize}
For example, the diamond lattice used for the Fu-Kane-Mele-Hubbard
model and the two-dimensional honeycomb lattice considered in
the previous subsection are unchanged under the $\mathcal{P}$ and
$\mathcal{R}$ transformations.  Then, one can find that the Hubbard
model Hamiltonian given in Eq.~\eqref{eq:hubbard}, tight-binding
Hamiltonian given in Eq.~\eqref{eq:TB}, and the spin-orbit
interaction given in Eq.~\eqref{eq:H_SO} are invariant under the
$\mathcal{P}$ and $\mathcal{R}$ transformations.

Starting with the model that is invariant under the $\mathcal{P}$
and $\mathcal{R}$ transformations, the effective theory
for the electrons in the orbitals of our interest should also respect these
symmetries.  In the effective theory, the interaction of the
next-nearest neighbor sites can be expressed as
\begin{align}
  H_{\rm NNN} = \sum_{\langle\langle i,j \rangle\rangle}
  \left( \ell_{a,ij} c_i^\dagger \sigma_a c_j + t_{ij} c_i^\dagger c_j \right),
\end{align}
where $\langle\langle i,j \rangle\rangle$ is a set of the next-nearest
neighbor sites.  (Here, we consider the effective theory containing
only the electrons in the unique orbital of our interest, and the index for the
electron orbital is omitted for the notational simplicity.)

Now, we discuss the properties of the coefficient $\ell_{a,ij}$ and
show that, with $\mathcal{P}$ and $\mathcal{R}$ symmetries,
$\vec{\ell}_{ij}$ is proportional to
$\vec{d}_{ij}^1\times\vec{d}_{ij}^2$.  To see this, we can use the
following relations:
\begin{align}
  \ell_{a,\hat{i}\hat{j}} c_{\hat{i}}^\dagger \sigma_a c_{\hat{j}}
  \xrightarrow{\mathcal{P}} &\,
  - \ell_{1,\hat{i}\hat{j}} c_{\hat{i}}^\dagger \sigma_1 c_{\hat{j}}
  - \ell_{2,\hat{i}\hat{j}} c_{\hat{i}}^\dagger \sigma_2 c_{\hat{j}}
  + \ell_{3,\hat{i}\hat{j}} c_{\hat{i}}^\dagger \sigma_3 c_{\hat{j}},
  \label{eq:P}
  \\
  \ell_{a,\hat{i}\hat{j}} c_{\hat{i}}^\dagger \sigma_a c_{\hat{j}}
  \xrightarrow{\mathcal{R}} &\,
  \ell_{1,\hat{i}\hat{j}} c_{\hat{j}}^\dagger \sigma_1 c_{\hat{i}}
  - \ell_{2,\hat{i}\hat{j}} c_{\hat{j}}^\dagger \sigma_2 c_{\hat{i}}
  - \ell_{3,\hat{i}\hat{j}} c_{\hat{j}}^\dagger \sigma_3 c_{\hat{i}}.
  \label{eq:R}
\end{align}
Eq.~\eqref{eq:P} results in $\ell_{1,ij}=\ell_{2,ij}=0$ while
Eq.~\eqref{eq:R} implies $\ell_{3,ji}=-\ell_{3,ij}$, and hence
we can find that
$\vec{\ell}_{ij}\propto\vec{d}_{ij}^1\times\vec{d}_{ij}^2$.

\section{Transformation of $\alpha$ matrix}
\label{sec:alpha}
\setcounter{equation}{0}

The chiral representation of $\alpha$ matrices are defined as
\begin{align}
	\alpha_i=\begin{pmatrix}
		\sigma_i & 0 \\ 0 &-\sigma_i
	\end{pmatrix},~~~
	\alpha_4=\begin{pmatrix}
		0 & -1 \\ -1 & 0
	\end{pmatrix},~~~
	\alpha_5=\begin{pmatrix}
		0 & -i \\ i & 0
	\end{pmatrix},
	\label{alpha_chiral}
\end{align}
where $\alpha_5=\alpha_1\alpha_2\alpha_3\alpha_4$. They satisfy the anti-commutation relation $\{\alpha_\mu,\alpha_\nu\}=2\delta_{\mu\nu}$.
Under the unitary transformation $\alpha_\mu\to\widetilde \alpha_\mu = U^\dagger \alpha_\mu U$, the anti-commutation relation remains intact. For some choice of $U$, the $\alpha$ matrices are exchanged. Examples are summarized in Table.~\ref{table:alpha},
where
\begin{align}
	U_1= \frac{1}{\sqrt 2}\begin{pmatrix}
		1 & 0 & 0 & -i \\
		0 & 1 & -i & 0 \\
		0 &-i &1 & 0\\
		-i& 0& 0& 1
	\end{pmatrix},~~
	U_2= \frac{1}{\sqrt 2}\begin{pmatrix}
		1 & 0 & 0 & -1 \\
		0 & 1 & 1 & 0 \\
		0 & -1 &1 & 0\\
		1 & 0& 0& 1
	\end{pmatrix},~~
	U_3= \frac{1}{\sqrt 2}\begin{pmatrix}
		1 & 0 & -i & 0 \\
		0 & 1 & 0 & i \\
		-i & 0 &1 & 0\\
		0 & i & 0 & 1
	\end{pmatrix}.
\end{align}
Note that they have the form of
\begin{align}
	U_i = \frac{1}{\sqrt 2}\begin{pmatrix}
		1 & -i\sigma_i \\
		-i\sigma_i & 1
	\end{pmatrix}.
\end{align}
for $i=1,2,3$. One can easily show that they yield
\begin{align}
	U_i^\dagger \alpha_j U_i =\begin{cases}
		\alpha_j & {\rm for}~~~i\neq j \\
		-\alpha_5 & {\rm for}~~~i=j
	\end{cases}.
\end{align}

The Dirac representation for the $\alpha$ matrices is given by
\begin{align}
	\alpha_i=\begin{pmatrix}
		0 & \sigma_i \\ \sigma_i & 0
	\end{pmatrix},~~~
	\alpha_4=\begin{pmatrix}
		1 & 0 \\ 0 & -1
	\end{pmatrix},~~~
	\alpha_5=\begin{pmatrix}
		0 & -i \\ i & 0
	\end{pmatrix},
	\label{alpha_Dirac}
\end{align}
The chiral and Dirac representations are related by the unitary transformation as
\begin{align}
	\alpha^{\rm (Dirac)}_\mu = U^\dagger \alpha^{\rm (chiral)}_\mu U,~~~~~~
	U= \frac{1}{\sqrt 2}\begin{pmatrix}
		1 & 1 \\
		-1 & 1
	\end{pmatrix}.
\end{align}

\begin{table}
\begin{center}
\begin{tabular}{ |c|c|c|c|c|c| } \hline
    ~& $\widetilde \alpha_1$ & $\widetilde \alpha_2$ & $\widetilde \alpha_3$ & $\widetilde \alpha_4$ & $\widetilde \alpha_5$\\ \hline
    $U_1$ & $\alpha_5$ & $\alpha_2$ & $\alpha_3$ & $\alpha_4$ & $-\alpha_1$\\ \hline
    $U_2$ & $\alpha_1$ & $\alpha_5$ & $\alpha_3$ & $\alpha_4$ & $-\alpha_2$\\ \hline
    $U_3$ & $\alpha_1$ & $\alpha_2$ & $\alpha_5$ & $\alpha_4$ & $-\alpha_3$\\ \hline
\end{tabular}
\caption{Transformation law of $\alpha$-matrices under the unitary transformation by $U_1,U_2$ and $U_3$.}
\label{table:alpha}
\end{center}
\end{table}

\section{Berry connection and topological term}
\label{sec:Berry}
\setcounter{equation}{0}

\subsection{Dimensional reduction of $(4+1)$-dimensional quantum Hall insulator}

In Sec.~\ref{sec:CM}, we derived $\theta$ using the Lagrangian formulation following Ref.~\cite{Sekine:2014xva}. On the other hand, $\theta$ can also be expressed in terms of the Berry connection~\cite{Qi:2008ew,Essin:2008rq}.

It is well known that the general $(2+1)$-dimensional quantum Hall insulator is characterized by the first Chern number $N_{\rm ch}^{(1)}$ in terms of the integration of the Berry connection over the Brillouin zone~\cite{Thouless:1982zz}. Its electromagnetic response is described by the action
\begin{align}
	S = \frac{N_{\rm ch}^{(1)}}{4\pi}\int dtd^2x\,\epsilon^{\mu\nu\rho}A_\mu \partial_\nu A_\rho.
\end{align}
Similarly, the $(4+1)$-dimensional quantum Hall insulator is characterized by the second Chern number $N_{\rm ch}^{(2)}$ and described by the action
\begin{align}
	S = \frac{N_{\rm ch}^{(2)}}{24\pi^2}\int dtd^4x\,\epsilon^{\mu\nu\rho\sigma\tau}A_\mu \partial_\nu A_\rho  \partial_\sigma A_\tau,
	\label{S5}
\end{align}
where
\begin{align}
	N_{\rm ch}^{(2)} = \frac{1}{32\pi^2}\int_{\rm BZ} d^4k\,\epsilon^{ijkl} {\rm Tr}\left[\mathcal F_{ij} \mathcal F_{kl}\right],
\end{align}
with
\begin{align}
	\mathcal F_{ij}\equiv \partial_i\mathcal A_j- \partial_j\mathcal A_i +i[\mathcal A_i,\mathcal A_j].
	\label{Nch2}
\end{align}
Here we used a shorthand notation like $\partial_i \equiv \partial/\partial k_i$ and so on ($k_4$ may be rather understood as $\varphi\equiv k_4+A_4$) and $\mathcal A_i$ denotes the Berry connection matrix in the momentum space given by
\begin{align}
	\mathcal A_i^{\alpha\beta} =-i \langle u_k^\alpha | \frac{\partial}{\partial k_i} | u_k^\beta\rangle.
\end{align}
with $|u_k^\alpha\rangle$ being the Bloch state with $\alpha$ representing the band index, and the trace in Eq.~(\ref{Nch2}) is taken over the occupied bands. Note that $N_{\rm ch}^{(2)}$ is expressed as
\begin{align}
	N_{\rm ch}^{(2)} = \frac{1}{2\pi}\int \frac{\partial \theta}{\partial \varphi} d\varphi,
\end{align}
where
\begin{align}
	\theta \equiv \frac{1}{4\pi}\int_{\rm BZ} d^3k\,\epsilon^{ijk}\,{\rm Tr}\left[ \mathcal A_i \partial_j \mathcal A_k+i\frac{2}{3} \mathcal A_i  \mathcal A_j \mathcal A_k\right].
	\label{theta}
\end{align}

Now let us perform a dimensional reduction. The action (\ref{S5}) is written as
\begin{align}
	S =\frac{1}{8\pi^2}\int dtd^3x\,\epsilon^{\mu\nu\rho\sigma}\frac{\partial\theta}{\partial \varphi} \partial_\mu \varphi A_\nu \partial_\rho A_\sigma
	=-\frac{1}{8\pi^2}\int dtd^3x\,\theta\,\epsilon^{\mu\nu\rho\sigma}\partial_\mu A_\nu \partial_\rho A_\sigma,
\end{align}
where we used $\partial_\mu\theta=(\partial\theta/\partial \varphi)\partial_\mu \varphi$. This is an action that describes the electromagnetic response of $(3+1)$-dimensional topological insulator.

\subsection{Hamiltonian expression of $\theta$}

Let us assume the four-band model whose (momentum space) Hamiltonian is given by
\begin{align}
	H= c_{k,\alpha}^\dagger \mathcal H_{\alpha\beta}  c_{k,\beta},~~~~~~~~\mathcal H = \sum_{\mu=1}^5 R_\mu(\vec k) \alpha_\mu,
	\label{Hamiltonian}
\end{align}
where $c_{k,\alpha}^\dagger$ and $c_{k,\alpha}$ with $\alpha=1$--$4$ denote the electron creation and annihilation operator with the wavenumber $k$ and $R_\mu$ are real coefficients. Here we take the Dirac representation for the $\alpha$ matrices (\ref{alpha_Dirac}).
The Hamiltonian (\ref{Hamiltonian}) is diagonalized by the unitary matrix $U$:
\begin{align}
	U = \begin{pmatrix}
		N_+(-R_1+iR_2) & N_+(-R_3+iR_5) & N_-(R_1-iR_2) &N_-(R_3-iR_5) \\
		N_+(R_3+iR_5) & N_+(-R_1-iR_2) & N_-(-R_3-iR_5) &N_-(R_1+iR_2) \\
		0 & N_+(R+R_4) & 0 & N_-(R-R_4) \\
		N_+(R+R_4) & 0 & N_-(R-R_4) & 0
	\end{pmatrix},
	\label{Umatrix}
\end{align}
where $N_\pm \equiv 1/\sqrt{2R(R\pm R_4)}$ and $R\equiv\sqrt{ \sum_{\mu=1-5}(R_\mu)^2}$. One finds
\begin{align}
	U^\dagger \mathcal H U =  {\rm diag}(-R,-R,R,R).
\end{align}
The lower two energy bands and upper two bands are degenerate and we assume that the lower bands are occupied and upper bands are empty. One can define the creation/annihilation operator in the diagonal basis through
\begin{align}
	d_{k,\alpha} \equiv U^\dagger_{\alpha\beta} c_{k,\beta},~~~~~~d_{k,\alpha}^\dagger \equiv c_{k,\beta}^\dagger U_{\beta\alpha}.
\end{align}
The Bloch state may be given by $|u_k^\alpha\rangle = d^\dagger_{k,\alpha}|0\rangle =c_{k,\beta}^\dagger U_{\beta\alpha}|0\rangle$. Thus the Berry connection is calculated as
\begin{align}
	\mathcal A_i^{\alpha\beta} =-i  \langle 0|  U^\dagger_{\alpha\gamma} c_{k,\gamma} \frac{\partial}{\partial k_i} (c_{k,\delta}^\dagger U_{\delta\beta}) |0\rangle
	= -i U^\dagger_{\alpha\gamma}\frac{\partial U_{\gamma\beta}}{\partial k_i}.
\end{align}
Note that $\mathcal A_i^{\alpha\beta}$ is a $2\times 2$ matrix since only the two low energy states are occupied. Substituting the concrete expression (\ref{Umatrix}), we obtain
\begin{align}
	\mathcal A_i = \sum_{a=1}^3 A_{ia} \sigma_a,
\end{align}
where
\begin{align}
	& A_{i1}=-N_+^2\left[ (R_1\partial_i R_5-R_5\partial_i R_1)+(R_3\partial_i R_2-R_2\partial_i R_3) \right], \\
	& A_{i2}=-N_+^2\left[ (R_3\partial_i R_1-R_1\partial_i R_3)+(R_5\partial_i R_2-R_2\partial_i R_5) \right], \\
	& A_{i3}=-N_+^2\left[ (R_1\partial_i R_2-R_2\partial_i R_1)+(R_5\partial_i R_3-R_3\partial_i R_5) \right].
\end{align}
Note that the term proportional to the unit matrix ${\bf 1}$ is canceled.

Using the trace formula ${\rm Tr}\left[\sigma_a\sigma_b\right]=2\delta_{ab}$ and ${\rm Tr}\left[\sigma_a\sigma_b\sigma_c\right]=2i\epsilon_{abc}$, the first and second terms of $\theta$ in (\ref{theta}) are calculated as
\begin{align}
	&\epsilon^{ijk}\,{\rm Tr}\left[\mathcal A_i \partial_j \mathcal A_k \right]
	=\frac{-3}{R^2 (R+R_4)^2}\epsilon^{\mu\nu\rho\sigma} R_\mu (\partial_x R_\nu)(\partial_y R_\rho)(\partial_z R_\sigma), \\
	&\epsilon^{ijk}\,{\rm Tr}\left[i\frac{2}{3}\mathcal A_i \mathcal A_j \mathcal A_k \right]
	=\frac{R^2-R_4^2}{R^3 (R+R_4)^3}\epsilon^{\mu\nu\rho\sigma}R_\mu (\partial_x R_\nu)(\partial_y R_\rho)(\partial_z R_\sigma),
\end{align}
where $\mu,\nu,\rho,\sigma = 1,2,3,5$. Note that terms proportional $\partial_j N_+$, $\partial_j \partial_k R_\mu$, $\partial_j R_1 \partial_k R_1$ and so on vanish when contracted by $\epsilon^{ijk}$. Thus we obtain the following expression for $\theta$,
\begin{align}
	\theta =-\frac{1}{4\pi}\int_{\rm BZ} d^3k \frac{2R+R_4}{R^3 (R+R_4)^2}\epsilon^{\mu\nu\rho\sigma}R_\mu (\partial_x R_\nu)(\partial_y R_\rho)(\partial_z R_\sigma).
	\label{theta_R}
\end{align}
This expression is consistent with \cite{Wang:2010}.\footnote{
	Note that the definition of $\alpha_3$ and $\alpha_5$ are reversed between us and Ref.~\cite{Wang:2010} and hence there appears an extra minus sign in the final expression (\ref{theta_R}).
}

\bibliography{bib}
\bibliographystyle{jhep}

\end{document}